\documentclass[aps,prl,twocolumn,nofootinbib]{revtex4-1} 

\usepackage{epsfig}
\usepackage{color}
\usepackage[dvipsnames]{xcolor}
\usepackage{float,amsmath}
\usepackage{graphicx}
\usepackage{hyperref}
\usepackage[utf8]{inputenc}

\newcommand{\tmop}[1]{\ensuremath{\operatorname{#1}}}

\newcommand{\bt}{{\mathbf{b}_\perp}}

\begin{document}

\title{Imprints of fluctuating proton shapes on flow in proton-lead collisions at the LHC}
\author{Heikki M\"antysaari}
\affiliation{Physics Department, Brookhaven National Laboratory, Upton, NY 11973, USA}

\author{Bj\"orn Schenke}
\affiliation{Physics Department, Brookhaven National Laboratory, Upton, NY 11973, USA}

\author{Chun Shen}
\affiliation{Physics Department, Brookhaven National Laboratory, Upton, NY 11973, USA}

\author{Prithwish Tribedy}
\affiliation{Physics Department, Brookhaven National Laboratory, Upton, NY 11973, USA}

\begin{abstract}
Results for particle production in $\sqrt{s}=5.02\,{\rm TeV}$ p+Pb collisions at the Large Hadron Collider within a combined classical Yang-Mills and relativistic viscous hydrodynamic calculation are presented. We emphasize the importance of sub-nucleon scale fluctuations in the proton projectile to describe the experimentally observed azimuthal harmonic coefficients $v_n$, demonstrating their sensitivity to the proton shape. We stress that the proton shape and its fluctuations are not free parameters in our calculations. Instead, they have been constrained using experimental data from HERA on exclusive vector meson production. Including temperature dependent shear and bulk viscosities, as well as UrQMD for the low temperature regime, we present results for mean transverse momenta, harmonic flow coefficients for charged hadrons and identified particles, as well as Hanbury-Brown-Twiss radii.
\end{abstract}

\maketitle


{\bf Introduction}
In heavy ion collisions strong final state interactions that are well described within the framework of relativistic hydrodynamics produce azimuthally anisotropic particle distributions that are strongly correlated with the initial event geometry \cite{Heinz:2013th,Gale:2013da,deSouza:2015ena,Song:2017wtw}.
Measurements of produced hadrons in collisions of protons and other small nuclei with heavy ions at the Relativistic Heavy Ion Collider (RHIC) and the Large Hadron Collider (LHC) have observed very similar azimuthal anisotropies as in heavy ion collisions \cite{Adare:2013piz,CMS:2012qk,Abelev:2012ola,Aad:2012gla} (also see the review \cite{Dusling:2015gta}). Naturally, it was suggested that these anisotropies in small systems could have the same origin as in heavy ion collisions, namely the hydrodynamic response of the produced medium to the initial shape of the interaction region in the transverse plane \cite{Bozek:2011if,Bozek:2012gr,Bozek:2013df,Bozek:2013uha,Bozek:2013ska,Qin:2013bha,Werner:2013ipa,Kozlov:2014fqa,Romatschke:2015gxa,Shen:2016zpp,Weller:2017tsr}.

It has also been shown that in high energy collisions of hadrons and nuclei, gluons are produced with intrinsic anisotropic azimuthal momentum distributions (for recent reviews see \cite{Dusling:2015gta,Schlichting:2016kjw}). These distributions have some features often attributed to hydrodynamic behavior \cite{Schenke:2016lrs,Dusling:2017dqg} but are uncorrelated with the global geometry of the interaction region \cite{Schenke:2015aqa,McLerran:2016snu}. Furthermore, it is not clear whether they will survive potential final state interactions. In this work we focus on high multiplicity events and assume that the initial state momentum correlations are fully erased by strong final state interactions. Thus anisotropies in final observables are enitrely due to the system's response to the initial geometry. We further assume that relativistic fluid dynamics is applicable in the presence of fairly large gradients (as present in small collision sytems such as p+Pb) and at times $\lesssim 1\,{\rm fm}$, something which has been extensively argued to be the case in the literature \cite{Chesler:2009cy,Wu:2011yd,Heller:2011ju,vanderSchee:2012qj,Casalderrey-Solana:2013aba,Kurkela:2015qoa,Keegan:2015avk,Chesler:2016ceu,Attems:2016tby,Romatschke:2016hle,Romatschke:2017vte}.

The initial state and early time evolution in this work is described by the IP-Glasma model \cite{Schenke:2012wb,Schenke:2012hg}. This model has been shown to lead to very good agreement with a wide range of experimental data \cite{Gale:2012rq,Schenke:2014zha}, 
and, in contrast to some other models, was shown to be compatible with data in an extensive Bayesian parameter estimation study \cite{Bernhard:2016tnd}. 
The only other models that are in good agreement with a similar amount of experimental data, including event-by-event distributions of flow harmonics \cite{Gale:2012rq,Niemi:2012aj,Schenke:2014zha,Aad:2013xma}, are the EKRT model \cite{Niemi:2015qia}, and a particular version of TRENTo \cite{Moreland:2014oya}.

The field energy-momentum tensor from the IP-Glasma model provides the initialization for the relativistic viscous hydrodynamic simulation \textsc{Music} \cite{Schenke:2010nt,Schenke:2010rr,Schenke:2011bn}. In the last step of the calculation we determine spatial and momentum distributions of all particles and feed them into the hadronic cascade UrQMD \cite{Bass:1998ca,Bleicher:1999xi} to obtain our final results.

Hydrodynamic evolution of IP-Glasma initial states in small systems has been explored before \cite{Bzdak:2013zma,Schenke:2014zha}. The most notable result from the analysis in \cite{Schenke:2014zha} was that the measured harmonic flow coefficients in p+Pb collisions were underestimated significantly. This stood in sharp contrast with calculations using the Monte-Carlo (MC) Glauber model for the initial state, which lead to much better agreement with experimental flow harmonics in small systems \cite{Bozek:2011if,Bozek:2012gr,Bozek:2013df,Bozek:2013uha,Bozek:2013ska,Qin:2013bha,Werner:2013ipa,Kozlov:2014fqa,Romatschke:2015gxa,Shen:2016zpp,Weller:2017tsr}. As anticipated in \cite{Schenke:2014zha} the reason for the disagreement with experimental data is the underestimation of subnucleonic fluctuations in the IP-Glasma model - the profile of a nucleon's color charge density was assumed to be Gaussian in the transverse plane (called 'round proton' in the following). In the Yang-Mills framework, the interaction region follows much more closely the shape of the smaller projectile than in MC Glauber implementations. This is why in p+A collisions one becomes sensitive to the assumed proton shape and its fluctuations, while in d/$^3$He/A+A collisions spatial eccentricities are dominated by nucleon size scale fluctuations.

Because fluctuations of the proton shape constitute an additional degree of freedom, their inclusion requires an additional external constraint. In the spirit of the IP-Glasma model, whose parameters are determined by HERA deeply inelastic scattering data \cite{Rezaeian:2012ji}, we employ additional data from HERA on diffractive $J/\Psi$ production \cite{Breitweg:1999jy,Chekanov:2002xi,Chekanov:2002rm,Aktas:2003zi,Aktas:2005xu,Alexa:2013xxa}. In a recent work by two of the authors the incoherent diffractive cross section for exclusive $J/\Psi$ production was used to constrain the degree of fluctuations of the gluon distribution in the proton within the IP-Glasma model \cite{Mantysaari:2016ykx,Mantysaari:2016jaz}. In this work we employ the model using three gluonic hot spots \cite{Schlichting:2014ipa} with parameters determined in \cite{Mantysaari:2016jaz}. Similar models were discussed recently in \cite{Albacete:2016pmp,Welsh:2016siu,Cepila:2016uku,Weller:2017tsr,Mantysaari:2017dwh,Moreland:2017kdx}.

We show in the following that strong final state interactions turn initial shape fluctuations of the proton's gluon distribution into observable final state particle correlations. The values of the resulting flow harmonics differ from those obtained for round protons by up to a factor of 5. This means that high energy p+A collisions provide unprecedented access to analyze the shape of the proton and its fluctuations. The constrained proton shape parameters from \cite{Mantysaari:2016ykx,Mantysaari:2016jaz} together with temperature dependent 
medium properties similar to those employed in a previous work \cite{Denicol:2015nhu} allow for a quantitative description of experimentally measured particle spectra, flow harmonics, and Hanbury-Brown-Twiss (HBT) radii in p+Pb collisions at $5.02\,{\rm TeV}$.

{\bf Fluctuating proton in the IP-Glasma initial state}
As discussed above, the original IP-Glasma model assumed a 2D-Gaussian spatial shape for all nucleons. This assumption lead to a dramatic underestimation of flow harmonics in p+A collisions \cite{Schenke:2014zha}. Here we introduce an additional substructure to all nucleons, using three hot spots whose positions in the transverse plane are Gaussian distributed with width $B_{qc}$. 
The density profile of each hot spot in the transverse plane is also assumed to be Gaussian
\begin{equation}
T_q(\bt) = \frac{1}{2\pi B_q} e^{-\bt^2/(2B_q)}\,,
\end{equation}
with width parameter $B_q$.
Other than this substitution for the round nucleon, the implementation of the IP-Glasma model is exactly as described in \cite{Schenke:2013dpa}.
The infrared regulator $m$ is determined together with $B_{qc}$ and $B_q$ from the analysis of HERA data in \cite{Mantysaari:2016jaz}. The values are $m=0.4\,{\rm GeV}$, $B_{qc}=3\,{\rm GeV}^{-2}$, and $B_q=0.3\,{\rm GeV}^{-2}$.

Because of the shorter lifetime in small systems, the initial viscous stress tensor from the classical Yang-Mills (CYM) simulation plays a more important role compared to heavy ion collisions. We thus include the full $T^{\mu\nu}_{\rm CYM}$ when initializing the hydrodynamic simulation as opposed to only extracting the energy density $\varepsilon$ and transverse flow velocities $u^\mu$ as done in previous works. Using the ideal equation of state $\varepsilon = 3 P$ of the classical Yang-Mills system we have
$$\pi^{\mu\nu}=T^{\mu\nu}_{\rm CYM}-\frac{4}{3}\varepsilon u^\mu u^\nu + \frac{\varepsilon}{3}g^{\mu\nu}\,.$$ We find the spatial dependence of components of $\pi^{\mu\nu}$ to be similar to the initial Navier-Stokes value when using a shear viscosity of $\eta/s=0.1$. We further include a correction for the initial value for the bulk component of the stress energy tensor to account for the different equation of state on the hydrodynamic side of the simulation, which is taken from lattice QCD \cite{Huovinen:2009yb}.
This matching is performed at the initial switching time $\tau_0$, which we vary from $\tau_0=0.2\,{\rm fm}$ to $\tau_0=0.4\,{\rm fm}$.

{\bf Hydrodynamics and hadronic cascade}
In the hydrodynamic simulation \textsc{Music} we use the same second order transport parameters as in \cite{Ryu:2015vwa}. We employ a temperature dependent shear viscosity to entropy density ratio with similar features as those found to describe rapidity dependent flow harmonics at RHIC energies in \cite{Denicol:2015nhu}. These features are a strong increase with decreasing temperature in the hadronic phase and no to little temperature dependence in the QGP phase. To be precise, we use the following parametrization
\begin{align}
(\eta/s)(T) &= (\eta/s)_{\rm min} + a (T_c-T)\theta(T_c-T) \notag\\ 
& ~~~~+ b (T-T_c)\theta(T-T_c)\,,
\end{align}
where $a=15$\,GeV$^{-1}$, $b=1$\,GeV$^{-1}$, and $(\eta/s)_{\rm min}=0.08$ at a temperature of $166\,{\rm MeV}$. We note that both the minimal value and the slope in the hadronic phase ($a$) are larger than what was used in \cite{Denicol:2015nhu}. The reason for needing larger viscosities to describe the p+Pb data could be that i) we are using a different initial state model and ii) we use parameters for the initial proton shape determined at $x\approx 10^{-3}$, while typical $x$ values in 5.02 TeV p+Pb collisions at midrapidity are $x\approx 2\cdot 10^{-4}$. Explicit small $x$ evolution via solution of the JIMWLK equation \cite{Jalilian-Marian:1997jx,Jalilian-Marian:1997gr,Iancu:2000hn} that we do not include here would lead to the reduction of eccentricities towards this lower $x$ \cite{Schlichting:2014ipa}, leading in turn to an extraction of smaller viscosity values. The inclusion of this energy dependence is numerically intensive but will be a very interesting project for the future.

We also employ an effective shear viscosity of $\eta/s=0.2$ for comparison. The bulk viscosity to entropy density ratio $\zeta/s$ used is similar to that in \cite{Denicol:2009am,Denicol:2015nhu,Ryu:2015vwa}, the difference being an exponential reduction of $\zeta/s$ at low temperatures to keep non-equilibrium corrections to the thermal distribution functions \cite{Rose:2014fba} under control. We note that in small systems the effect of bulk viscosity on the evolution is essential to reproduce the $\langle p_T \rangle$ of identified hadrons, even more so than in heavy ion collisions \cite{Ryu:2015vwa}. A potential problem is that bulk viscosity in combination with the large expansion rates in p+Pb collisions can lead to negative effective pressures. We do not stop the hydrodynamic evolution when such negative values appear, whose existence has lead to the discussion of the potential phenomenon of cavitation \cite{Rajagopal:2009yw,Bhatt:2011kr,Habich:2014tpa,Denicol:2015bpa,Sanches:2015vra}. We note that anisotropic hydroynamics \cite{Nopoush:2014pfa,Alqahtani:2015qja} avoids (or at least reduces in a more general implementation) negative pressures by resumming certain viscous corrections. 

At a temperature of $T_{\rm switch}=155\,{\rm MeV}$ we determine the switching surface and sample particles that then propagate in the hadronic cascade model UrQMD. Particle momentum distributions are sampled first using the probability distribution function obtained from integrating the Cooper-Frye formula over the whole hyper-surface. Then the spatial positions of the sampled particles are determined from the differential Cooper-Frye formula. This sampling procedure introduces the least distortion in the momentum distribution of particle samples when the Cooper-Frye formula takes on negative values in certain regions of the hypersurface \cite{Shen:2014vra}. 

We have checked by explicit calculation that for charged hadrons the effect of rescattering in UrQMD is negligible. Proton spectra and $v_n(p_T)$ are slighty blue shifted because of the additional hadronic scatterings \cite{Shen:2016zpp}.

\begin{figure}[tb]
  \includegraphics[width=0.5\textwidth]{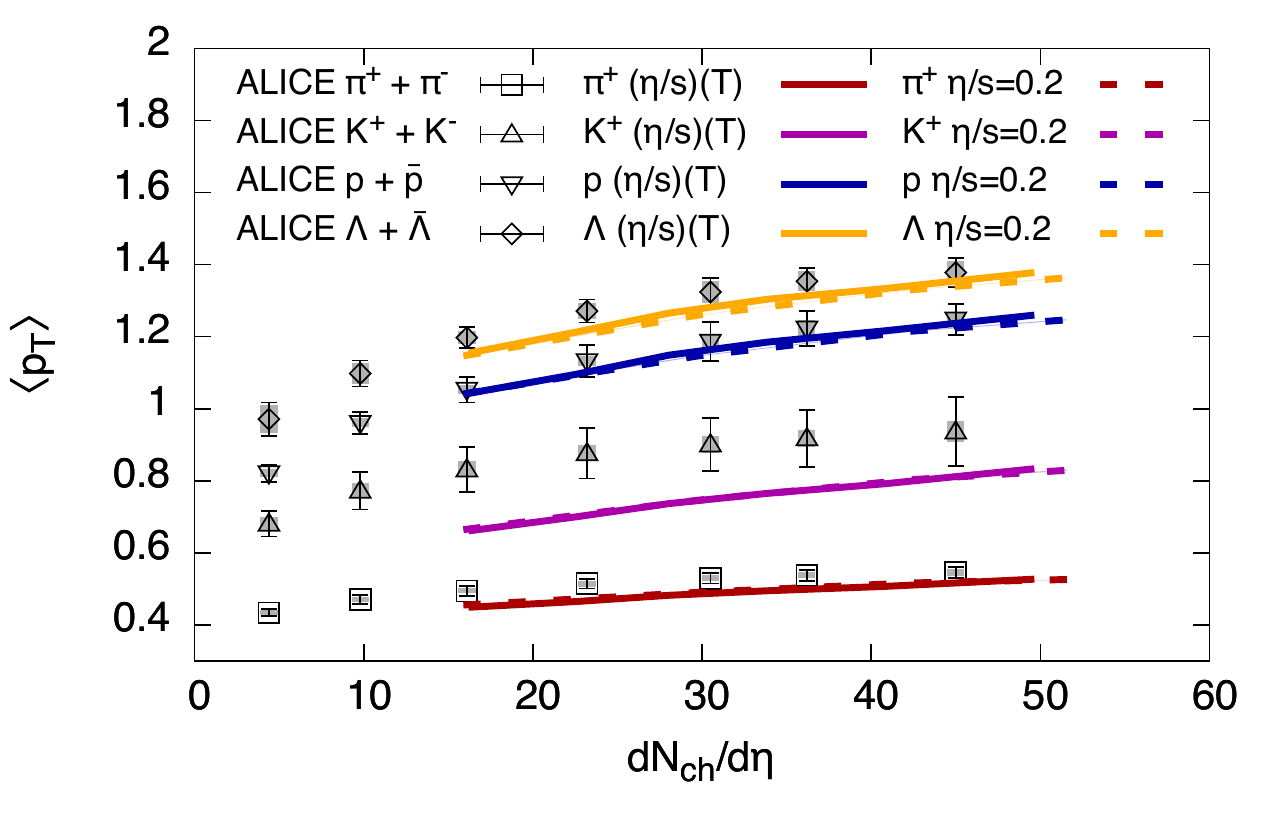}
  \caption{The mean transverse momentum $\langle p_T \rangle$ of identified particles as a function of the number of charged hadrons per pseudo-rapidity interval around mid-rapidity compared to experimental data from the ALICE collaboration \cite{Abelev:2013haa}. \label{fig:pt-pi-ALICE}}
\end{figure}

{\bf Results}
We begin by presenting results for the average transverse momentum $\langle p_T \rangle$ of identified particles as a function of charged particle multiplicity in Fig.\,\ref{fig:pt-pi-ALICE}. Using both temperature dependent $\eta/s$ and $\zeta/s$ and a switching time of $\tau_0=0.4\,{\rm fm}$ we find good agreement with experimental data from the ALICE collaboration for charged pions, protons, and $\Lambda$'s. The $\langle p_T \rangle$ of charged kaons is underestimated. We note that as discussed for heavy ion collisions in \cite{Ryu:2015vwa}, the inclusion of bulk viscosity is essential in order not to overestimate $\langle p_T \rangle$. Without bulk viscosity, the pion $\langle p_T \rangle$ is overestimated the most, by approximately $50\%$. The effect of using the constant effective $\eta/s$ is weak as is the effect of a smaller switching time $\tau_0=0.2\,{\rm fm}$, which is not shown here.

Having established the agreement with measured transverse momentum spectra, which is almost entirely determined by $\langle p_T \rangle$, we now present results for $v_n$ from two-particle correlations.  
To compute $v_n \{ 2 \}$ using particle samples from UrQMD, we first construct
the flow vector $Q_n = \sum_i w_i e^{i n \phi_i}$,
where the sum $i$ runs over all particles of interest with $0.3\,{\rm GeV} < p_T < 3\,{\rm GeV}$ (when comparing to CMS results), and the weights are set to $w_i = 1$. 
The two particle cumulant $v_n \{ 2 \}$ is then computed as
\begin{equation}
  v_n \{ 2 \} = \frac{1}{\langle N (N - 1) \rangle_{\rm ev}} (\langle
  \tmop{Re} \{ Q_n Q_n^{\ast} \} - N \rangle_{\rm ev})\,, \label{eq:vn}
\end{equation}
where $N$ is the number of particles included in the calculation of $Q_n$ and $\langle \cdot \rangle_{\rm ev}$ is the average over events. In practice, we sample the hypersurface from each hydrodynamic event 5000 times and run UrQMD for each of these particle configurations. For the evaluation of $v_n\{2\}$ we combine the UrQMD output of all 5000 runs to collect enough statistics and suppress short range correlations from e.g. resonance decays.
The latter effect is desired because the measurement uses a large pseudo-rapidity gap of $|\Delta\eta|>2$ between the two particles, also eliminating short range correlations.
\begin{figure}[tb]
  \includegraphics[width=0.5\textwidth]{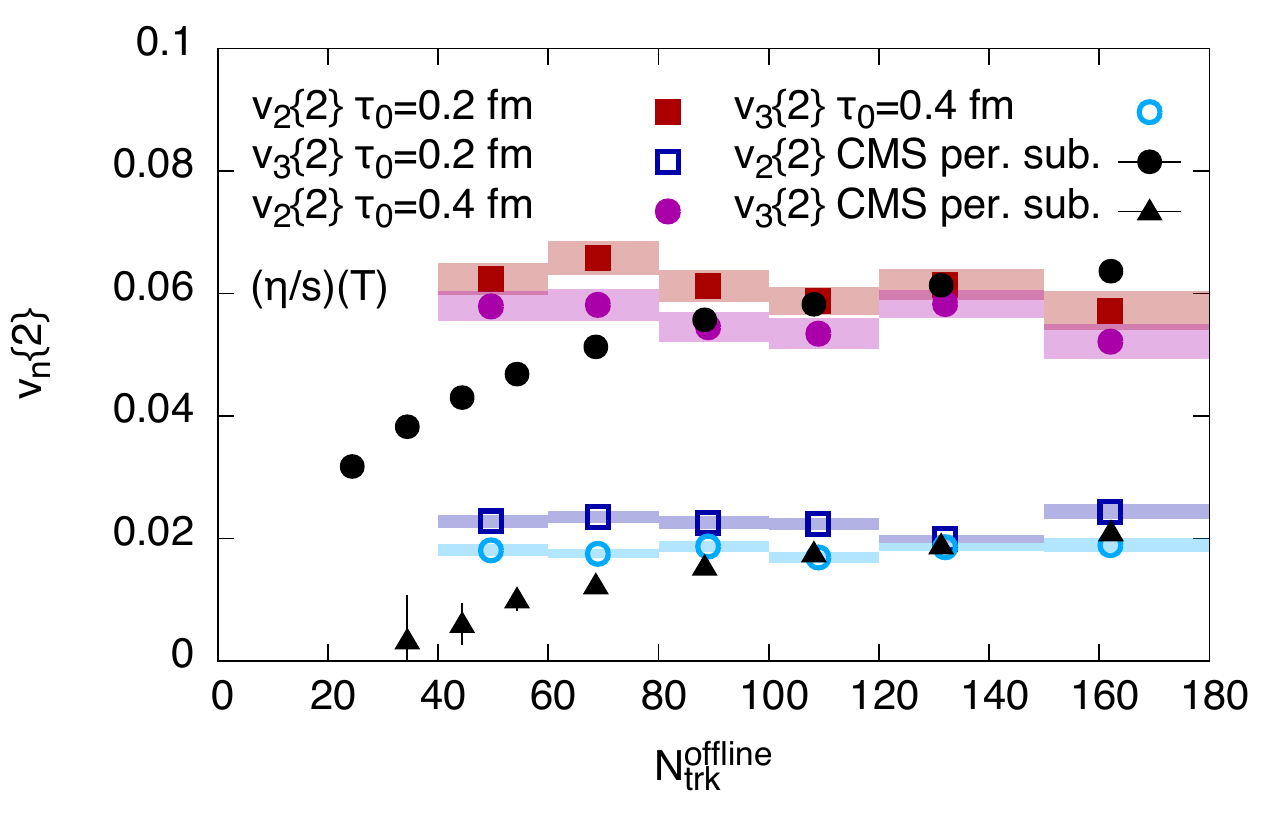}
  \caption{The second and third harmonic $v_2\{2\}$ and $v_3\{2\}$ of charged hadrons as a function of the number of tracks (as defined by the CMS collaboration) for the temperature dependent $\eta/s$ using $\tau_0=0.2\,{\rm fm}$ (squares) and $\tau_0=0.4\,{\rm fm}$ (circles). We compare to experimental data from the CMS Collaboration \cite{Chatrchyan:2013nka} with peripheral events subtracted. \label{fig:vn-pi}}
\end{figure}

\begin{figure}[tb]
  \includegraphics[width=0.5\textwidth]{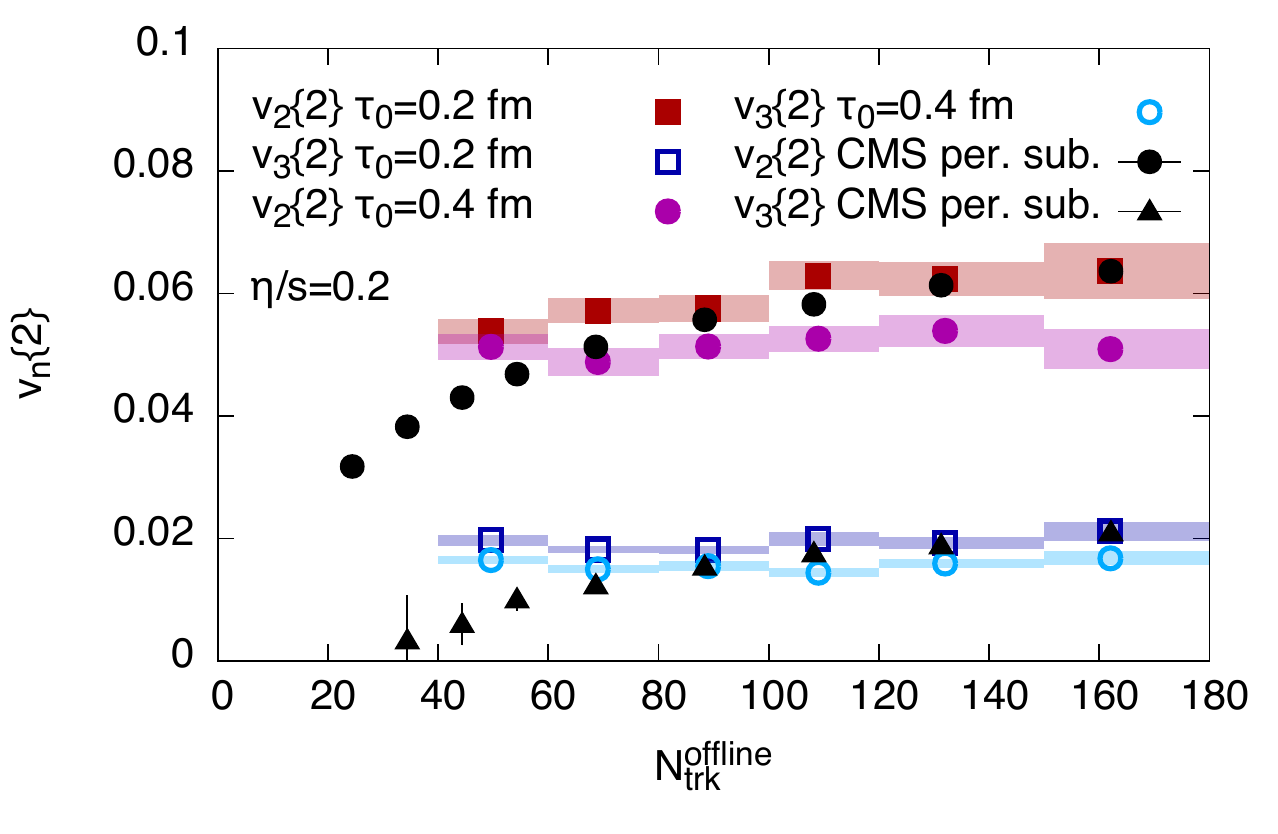}
  \caption{Same as Fig.~\ref{fig:vn-pi}, but for the effective $\eta/s=0.2$. \label{fig:vn-pi-02fm-points}}
\end{figure}

In Fig.\,\ref{fig:vn-pi} we see that above a multiplicity of $N_{\rm trk}^{\rm offline} \simeq 80 = 2\langle N_{\rm trk}^{\rm offline}\rangle $, the experimental $v_n\{2\}$ are well reproduced by our calculation, considering the uncertainty from the initial switching time, which we vary from $\tau_0=0.2\,{\rm fm}$ to $\tau_0=0.4\,{\rm fm}$.\footnote{To represent our results as a function of the experimentally determined $N_{\rm trk}^{\rm offline}$, we rescaled our computed multiplicities by a factor $\langle N_{\rm trk}^{\rm offline}\rangle/\langle d N_{\rm ch}/d\eta \rangle$.} Below $N_{\rm trk}^{\rm offline} \simeq 80 = 2\langle N_{\rm trk}^{\rm offline}\rangle$ the $v_n\{2\}$ are overestimated. This could indicate the failure of the hydrodynamic framework at low multiplicity. However, the experimental procedure to subtract correlations from peripheral events may cause a too fast decrease of $v_n\{2\}$ with decreasing multiplicity.

We repeat the same comparison in Fig.\,\ref{fig:vn-pi-02fm-points} using the constant effective $\eta/s=0.2$. 
We do not find a large difference to results with temperature dependent $\eta/s$ as long as the temperature does not rise too rapidly in the QGP phase. A more rapid rise ($b>2\,{\rm GeV}^{-1}$) was already ruled out in \cite{Denicol:2015nhu} and is not considered here.
As was already visible in Fig.\,\ref{fig:vn-pi}, the earlier switching to hydrodynamics leads to an increased anisotropic flow. This dependence on the switching time indicates that a more careful treatment of the non-equilibrium early time evolution is required. 

\begin{figure}[tb]
  \includegraphics[width=0.5\textwidth]{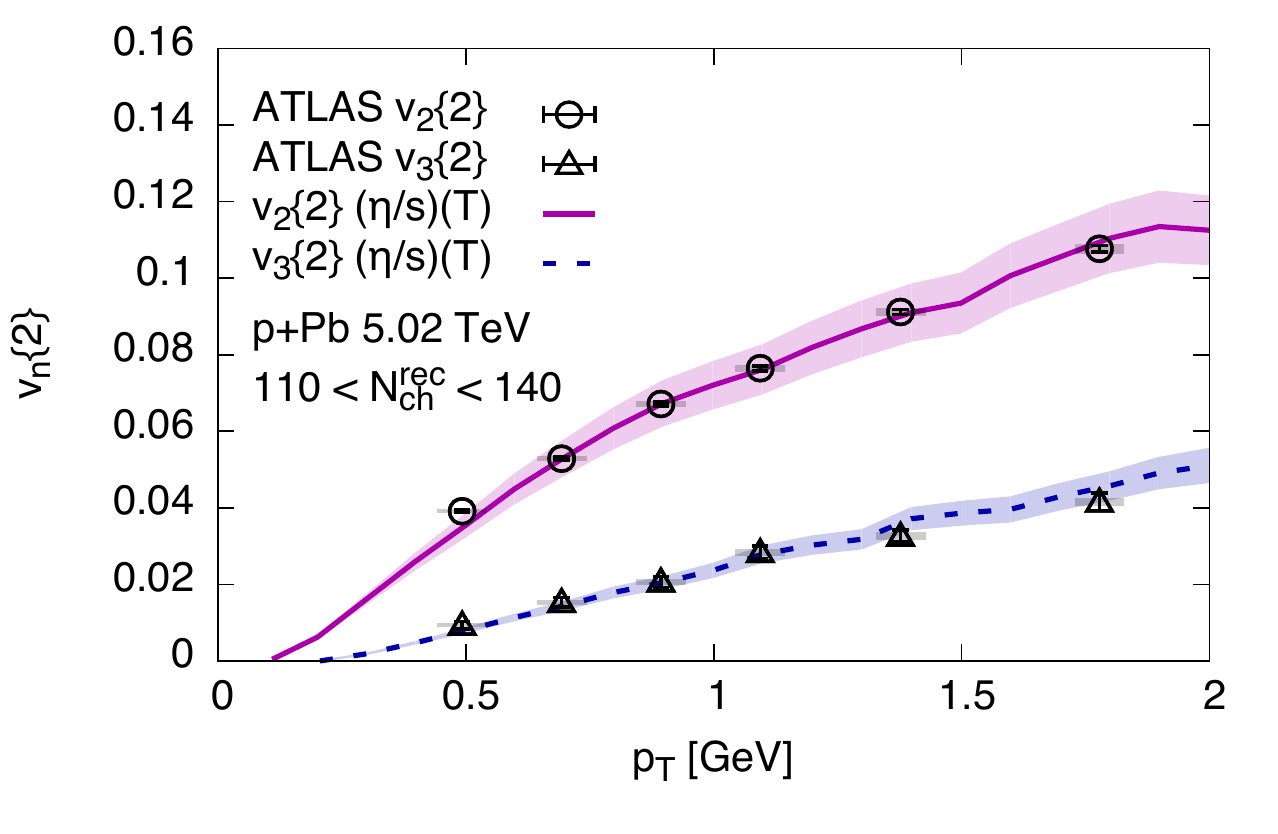}
  \caption{The second and third harmonic $v_2\{2\}$ and $v_3\{2\}$ of charged hadrons as a function of transverse momentum $p_T$ in one centrality class corresponding to $110 \leq N_{\rm ch}^{\rm rec} < 140$ (as defined by the ATLAS collaboration) for temperature dependent $\eta/s$. We compare to experimental data from the ATLAS Collaboration \cite{Aad:2014lta}. \label{fig:vn-pi-pt}}
\end{figure}

In Fig.\,\ref{fig:vn-pi-pt} we present our results for the transverse momentum dependent $v_2\{2\}(p_T)$ and $v_3\{2\}(p_T)$ in one multiplicity class and compare to experimental data from the ATLAS collaboration \cite{Aad:2014lta}. Here, the reference bin for the second particle is $1\,{\rm GeV}<p_T<3\,{\rm GeV}$. We find excellent agreement with the experimental data for the temperature dependent $\eta/s$. Results for the effective $\eta/s=0.2$ are slightly lower (not shown) but agree within statistical errors.

\begin{figure}[tb]
  \includegraphics[width=0.5\textwidth]{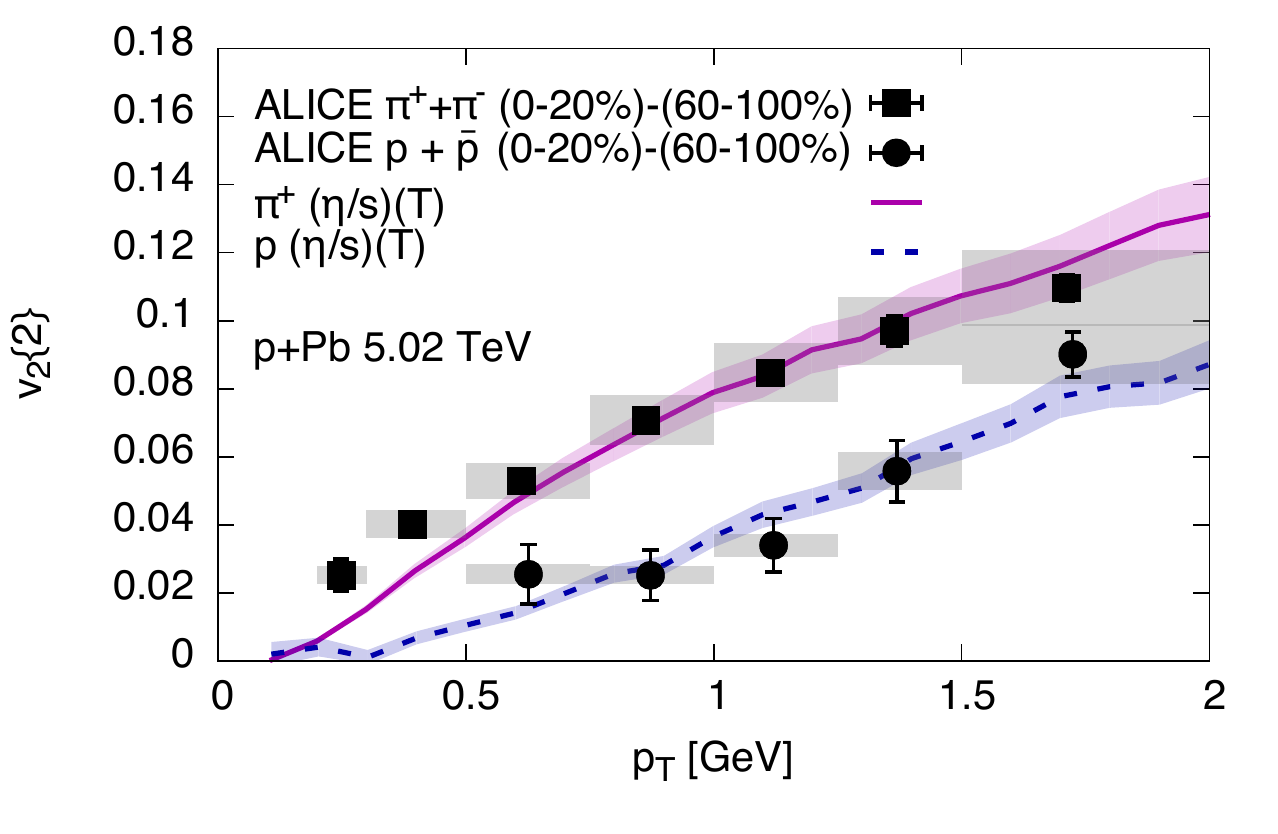}
  \caption{The second harmonic $v_2\{2\}$ of protons and pions as a function of transverse momentum $p_T$ for $0-20\%$ central events corresponding to $ d N_{\rm ch}/d\eta > 2 \langle d N_{\rm ch}/d\eta \rangle$.  We compare to experimental data of $v_2(2{\rm PC})$ from the ALICE Collaboration for 0-20\% centrality with correlations from 60-100\% centrality subtracted \cite{ABELEV:2013wsa}. \label{fig:vn-pPb-fluc-0-20pct-ID}}
\end{figure}

In Fig.\,\ref{fig:vn-pPb-fluc-0-20pct-ID} we present the transverse momentum dependent $v_2$ of identified pions and protons in $0-20\%$ central events. We compare with experimental data from the ALICE experiment \cite{ABELEV:2013wsa}, where the reference $p_T$ bin is $0.3\,{\rm GeV} < p_T < 4\,{\rm GeV}$. The characteristic mass dependence of $v_2(p_T)$ is well reproduced for $p_T<2\,{\rm GeV}$ and when comparing to results with correlations from 60-100\% central events subtracted. 

\begin{figure}[tb]
  \includegraphics[width=0.5\textwidth]{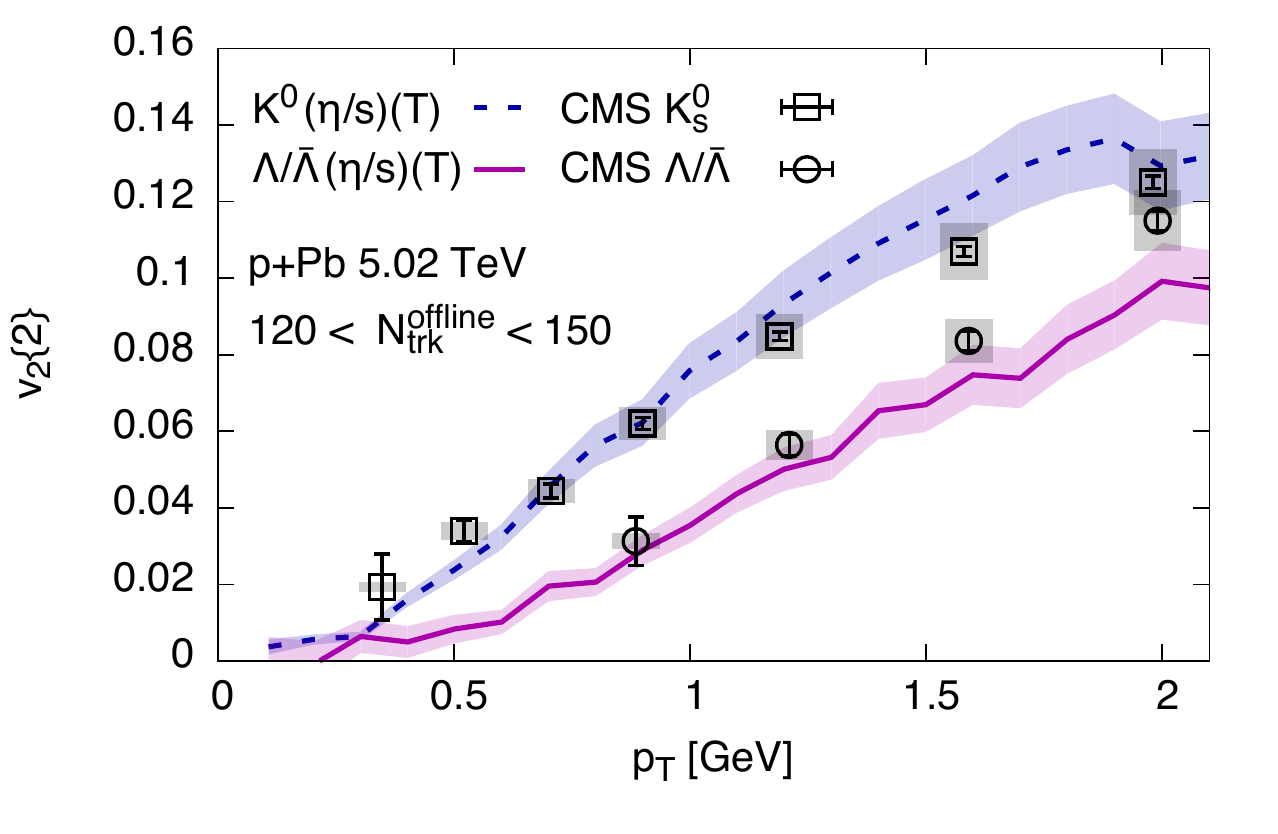}
  \caption{The second harmonic $v_2\{2\}$ of $\Lambda$ baryons and $K^0_S$ as a function of transverse momentum $p_T$ for $120 < N_{\rm trk}^{\rm offline} < 150$.  We compare to experimental data from the CMS Collaboration \cite{Khachatryan:2014jra}. \label{fig:vn-pPb-fluc-3-375-Lambda}}
\end{figure}

In Fig.\,\ref{fig:vn-pPb-fluc-3-375-Lambda} we present the $v_2(p_T)$ of $\Lambda$ baryons and $K^0_S$. Again, a clear mass dependence is visible. Agreement with experimental data from the CMS collaboration \cite{Khachatryan:2014jra} is very good. 
The results in Figs.\,\ref{fig:vn-pPb-fluc-0-20pct-ID} and \ref{fig:vn-pPb-fluc-3-375-Lambda} demonstrate that the mass ordering of $v_2$, while also possible to obtain from initial state momentum correlations and string fragmentation \cite{Schenke:2016lrs}, is quantitatively reproduced in our hydrodynamic simulations of p+Pb collisions.

Finally we present results for the HBT radii $R_{\rm long}$, $R_{\rm side}$, and $R_{\rm out}$, using both UrQMD results that include hadronic rescattering, and a calculation, independent of UrQMD, that includes only resonance decays, similar to \textsc{Therminator} \cite{Chojnacki:2011hb}.
For a given range in particle pair momentum $\mathbf{k}_T=(\mathbf{p}_T^1+\mathbf{p}_T^2)/2$, and under the assumption of a plane wave superposition for the pion wave function, the two-particle correlation function can be written as
\begin{eqnarray}
C(\mathbf{q}) &=& 1 + \frac{\frac{1}{\langle N_\mathrm{pair} \rangle} \langle \sum_{ij} \cos (q_{ij} \cdot x_{ij}) \rangle}{\frac{1}{\langle N_\mathrm{mixpair} \rangle} \langle N_\mathrm{mixpair}(q) \rangle},
\label{eq:HBT}
\end{eqnarray}
where $ij$ runs over all particle pairs $(i, j)$ within the $k_T=|\mathbf{k}_T|$ bin, and $q^\mu_{ij} = p^\mu_i - p^\mu_j$ and $x^\mu_{ij} = x_i^\mu - x_j^\mu$, with $x^\mu_i$ the space-time four vector of the last interaction point of particle $i$. $\langle N_{\rm pair}\rangle$ is the average total number of all pairs from single events, $\langle N_\mathrm{mixpair} \rangle$ the average total number of pairs from different (hydrodynamic) events, and $\langle N_\mathrm{mixpair}(q) \rangle$ is the average number of pairs from different events in the considered $q$-range.
For this analysis we also perform many UrQMD runs for a single hydrodynamic event. This results in values of $\langle N_{\rm pair}\rangle$ and  $\langle N_{\rm mixpair}\rangle$ of the order of $10^8$ per $k_T$ bin. 

\begin{figure}[tb]
  \includegraphics[width=0.5\textwidth]{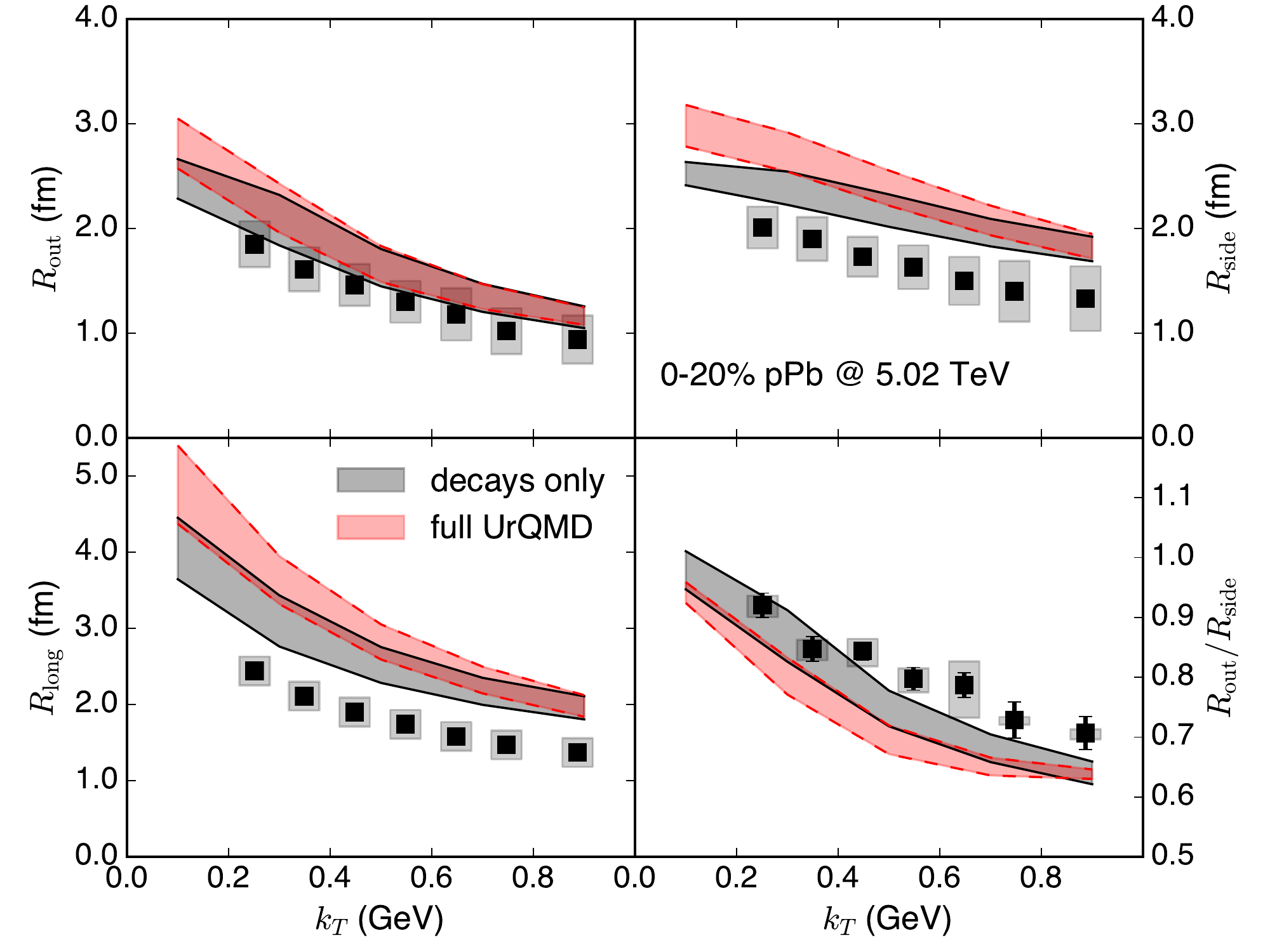}
  \caption{HBT radii extracted using a Gaussian fit to the correlation function (\ref{eq:HBT}) in the case of resonance decays only (band with solid lines) and UrQMD with interactions (band with dashed lines). We compare to experimental data from the \nobreak{ALICE} Collaboration \cite{Adam:2015pya}. See text for details. \label{fig:HBT_radii_pPb_Tdep}}
\end{figure}

The constructed two-pion HBT correlation function is fitted to the Pratt-Bertsch parameterization in the longitudinally co-moving system (LCMS) \cite{Pratt:1986cc,Bertsch:1988db}. We note that the shape of $C(\mathbf{q})$ is rather different from a Gaussian once resonance decay contributions are included. This non-Gaussianity results in different HBT radii depending on the fit range \cite{Plumberg:2016sig}. This is illustrated by the shaded bands in Fig.\,\ref{fig:HBT_radii_pPb_Tdep}, where we show $R_{\rm long}$, $R_{\rm side}$, and $R_{\rm out}$ along with the ratio $R_{\rm out}/R_{\rm side}$ as functions of $k_T$ for temperature dependent $\eta/s$. For the radii, the upper end of the bands results from fitting to the smallest momentum range $0.05\,{\rm GeV} < |\mathbf{q}| < 0.08\,{\rm GeV}$, the lower end to the largest, $0.05\,{\rm GeV} < |\mathbf{q}| < 0.2\,{\rm GeV}$. The systematic error quoted by the ALICE collaboration includes the variation of the upper limit of the fit range from $0.3\,{\rm GeV}$ to $1.1\,{\rm GeV}$. The limited statistics at large $|\mathbf{q}|$ in our calculation constrains the possible fit range to smaller values. 

We find that for $R_{\rm out}$ the lower end of both bands (dashed lines for the case of UrQMD including scattering in the later hadronic phase, solid lines for the case of only including decays) are compatible with the experimental data from the ALICE Collaboration \cite{Adam:2015pya}. The other two radii are overestimated by both calculations, resulting also in a slight underestimation of the ratio $R_{\rm out}/R_{\rm side}$, particularly in the case of using UrQMD. A smoother initial state (e.g. from small $x$ evolution) and reduced bulk viscosity could overcome this tension. Results with the constant effective $\eta/s$ are almost identical to the ones shown.

These results are comparable to previously published HBT radii from calculations using various other initial state models \cite{Bozek:2013df,Shapoval:2013jca,Romatschke:2015gxa}. In our calculation, the smaller (compared to e.g. the MC-Glauber model \cite{Bozek:2013df}) initial size of the system is overcome by a more explosive expansion and the accompanying extended lifetime due to entropy production from bulk viscous effects, leading to HBT radii being slightly larger than the experimental measurement. 

{\bf Discussion}
\emph{Early-time non-equilibrium phase:}
Given the shorter life-time compared to heavy ion collisions, small systems such as p+Pb collisions are more sensitive to the early time non-equilibrium stage of the evolution. We have performed calculations that initialized the hydrodynamic simulation directly with the energy momentum tensor from the classical Yang-Mills IP-Glasma calculation at times $\tau_0=0.2\,{\rm fm}$ and $0.4\,{\rm fm}$. All deviations from equilibrium and isotropy are thereby absorbed into viscous corrections on the hydrodynamic side of the simulation.  When not including the initial viscous stress tensor, values of integrated $v_n$ increase by up to 50\% (for $\tau_0=0.4\,{\rm fm}$, $\eta/s=0.2$ for $3-3.75\langle N_{\rm trk}^{\rm offline}\rangle$) underlining the importance of early times in p+Pb collisions.
So it is desirable to improve on this prescription in the future by including a realistic dynamic description of the non-equilibrium stage, which will smoothly interpolate between the IP-Glasma initial state and the viscous hydrodynamic description, without requiring very large initial viscous shear stress tensors in the hydrodynamic stage. A first attempt at this, using an effective kinetic theory, was presented in \cite{Keegan:2016cpi}.

\emph{Off-equilibrium corrections:}
Surprisingly, effects of off-equilibrium corrections to the thermal distribution functions largely cancel in final results for $p_T$-spectra or $v_n(p_T)$, however, studying the distribtution of corrections relative to the thermal distribution from all freeze-out surface cells, we find a significant share (10\% for $p_T\sim 0.45\,{\rm GeV}$, 25\% for $p_T\sim 1\,{\rm GeV}$, 45\% for $p_T\sim 1.5\,{\rm GeV}$ for pions) of large (shear) corrections ($|\delta f|/f \gtrsim 100\%$). This demonstrates that our results are plagued by large viscous corrections, in particular for $p_T\gtrsim 1\,{\rm GeV}$. Nevertheless, $p_T$-integrated quantities are dominated by low $p_T$ contributions and less sensitive to these problems.

\emph{Boost-invariance:} The IP-Glasma model is by construction boost invariant. It will be very interesting to use an extension of this model discussed in \cite{Schenke:2016ksl} to study p+Pb collisions within 3+1 dimensional hydrodynamic simulations. We note, however, that earlier work using a MC-Glauber initial state has found that at $\sqrt{s}=5.02\,{\rm TeV}$ differences of midrapidity $v_n$ between full 3+1 dimensional and boost invariant 2+1D simulations are negligible \cite{Shen:2016zpp}.

{\bf Conclusions}
We have presented results of particle production, anisotropic flow, and HBT radii in $\sqrt{s}=5.02\,{\rm TeV}$ proton lead collisions at the LHC from a hydrodynamic framework.
We use an initial state model that is based on an effective theory of QCD and compatible with a wide range of heavy ion data (IP-Glasma). This is coupled to a relativistic viscous hydrodynamic simulation (\textsc{Music}), which finally is connected to a microscopic hadronic afterburner (UrQMD). 

The main new ingredient in this work is the inclusion of subnucleonic fluctuations. These were independently shown to be highly relevant by comparison of the IP-Glasma calculation of diffractive $J/\Psi$ production with HERA data, which also allowed us to constrain the degree of subnucleonic fluctuations quantitatively. 

We have demonstrated that including subnucleonic fluctuations increases flow harmonics in p+Pb collisions by approximately a factor of 5 compared to previous calculations using round nucleons \cite{Schenke:2014zha}. Furthermore, we have shown that the multiplicity dependence of transverse momentum spectra, $v_2$ and $v_3$, as well as the transverse momentum dependence of the charged hadron and identified particle flow harmonics is well described within our framework, at least for multiplicities above $\sim 2$ times the average multiplicity. For lower multiplicities, we expect initial state momentum correlations (see \cite{Dusling:2015gta,Schlichting:2016kjw}) to become more dominant, and more sensitivity to the experimental method for removing jet-like correlations (new methods show little decrease of e.g.~$v_2\{4\}$ with decreasing multiplicity \cite{Jia:2017hbm}).

With the exception of $R_{\rm out}$, HBT radii are overestimated. However, a direct comparison with experimental data is difficult because of the non-Gaussian shape of the experimental and theoretical correlation functions, and limited statistics of the calculation at large momentum differences.

Future analyses with much higher statistics could be used to constrain the fine details of the subnucleonic structure in the initial state, in particular when higher order cumulants and other more complex multi-particle correlators can be studied. With that we could get unprecedented access to the fluctuating shape of protons and its energy dependence via proton-nucleus collisions.

{\bf Acknowledgments} 
We thank Christopher Plumberg for useful discussions about HBT correlations.
The authors are supported under DOE Contract No. DE-SC0012704. This research used resources of the National Energy Research Scientific Computing Center, which is supported by the Office of Science of the U.S. Department of Energy under Contract No. DE-AC02-05CH11231. BPS acknowledges a DOE Office of Science Early Career Award.

\vspace{-0.5cm}
\bibliography{spires}

\begin{thebibliography}{97}%
\makeatletter
\providecommand \@ifxundefined [1]{%
 \@ifx{#1\undefined}
}%
\providecommand \@ifnum [1]{%
 \ifnum #1\expandafter \@firstoftwo
 \else \expandafter \@secondoftwo
 \fi
}%
\providecommand \@ifx [1]{%
 \ifx #1\expandafter \@firstoftwo
 \else \expandafter \@secondoftwo
 \fi
}%
\providecommand \natexlab [1]{#1}%
\providecommand \enquote  [1]{``#1''}%
\providecommand \bibnamefont  [1]{#1}%
\providecommand \bibfnamefont [1]{#1}%
\providecommand \citenamefont [1]{#1}%
\providecommand \href@noop [0]{\@secondoftwo}%
\providecommand \href [0]{\begingroup \@sanitize@url \@href}%
\providecommand \@href[1]{\@@startlink{#1}\@@href}%
\providecommand \@@href[1]{\endgroup#1\@@endlink}%
\providecommand \@sanitize@url [0]{\catcode `\\12\catcode `\$12\catcode
  `\&12\catcode `\#12\catcode `\^12\catcode `\_12\catcode `\%12\relax}%
\providecommand \@@startlink[1]{}%
\providecommand \@@endlink[0]{}%
\providecommand \url  [0]{\begingroup\@sanitize@url \@url }%
\providecommand \@url [1]{\endgroup\@href {#1}{\urlprefix }}%
\providecommand \urlprefix  [0]{URL }%
\providecommand \Eprint [0]{\href }%
\providecommand \doibase [0]{http://dx.doi.org/}%
\providecommand \selectlanguage [0]{\@gobble}%
\providecommand \bibinfo  [0]{\@secondoftwo}%
\providecommand \bibfield  [0]{\@secondoftwo}%
\providecommand \translation [1]{[#1]}%
\providecommand \BibitemOpen [0]{}%
\providecommand \bibitemStop [0]{}%
\providecommand \bibitemNoStop [0]{.\EOS\space}%
\providecommand \EOS [0]{\spacefactor3000\relax}%
\providecommand \BibitemShut  [1]{\csname bibitem#1\endcsname}%
\let\auto@bib@innerbib\@empty
\bibitem [{\citenamefont {Heinz}\ and\ \citenamefont
  {Snellings}(2013)}]{Heinz:2013th}%
  \BibitemOpen
  \bibfield  {author} {\bibinfo {author} {\bibfnamefont {U.}~\bibnamefont
  {Heinz}}\ and\ \bibinfo {author} {\bibfnamefont {R.}~\bibnamefont
  {Snellings}},\ }\href {\doibase 10.1146/annurev-nucl-102212-170540}
  {\bibfield  {journal} {\bibinfo  {journal} {Ann. Rev. Nucl. Part. Sci.}\
  }\textbf {\bibinfo {volume} {63}},\ \bibinfo {pages} {123} (\bibinfo {year}
  {2013})},\ \Eprint {http://arxiv.org/abs/1301.2826} {arXiv:1301.2826
  [nucl-th]} \BibitemShut {NoStop}%
\bibitem [{\citenamefont {Gale}\ \emph
  {et~al.}(2013{\natexlab{a}})\citenamefont {Gale}, \citenamefont {Jeon},\ and\
  \citenamefont {Schenke}}]{Gale:2013da}%
  \BibitemOpen
  \bibfield  {author} {\bibinfo {author} {\bibfnamefont {C.}~\bibnamefont
  {Gale}}, \bibinfo {author} {\bibfnamefont {S.}~\bibnamefont {Jeon}}, \ and\
  \bibinfo {author} {\bibfnamefont {B.}~\bibnamefont {Schenke}},\ }\href@noop
  {} {\bibfield  {journal} {\bibinfo  {journal} {Int. J. of Mod. Phys. A, Vol.
  28,}\ }\textbf {\bibinfo {volume} {1340011}} (\bibinfo {year}
  {2013}{\natexlab{a}})},\ \Eprint {http://arxiv.org/abs/1301.5893}
  {arXiv:1301.5893 [nucl-th]} \BibitemShut {NoStop}%
\bibitem [{\citenamefont {de~Souza}\ \emph {et~al.}(2015)\citenamefont
  {de~Souza}, \citenamefont {Koide},\ and\ \citenamefont
  {Kodama}}]{deSouza:2015ena}%
  \BibitemOpen
  \bibfield  {author} {\bibinfo {author} {\bibfnamefont {R.~D.}\ \bibnamefont
  {de~Souza}}, \bibinfo {author} {\bibfnamefont {T.}~\bibnamefont {Koide}}, \
  and\ \bibinfo {author} {\bibfnamefont {T.}~\bibnamefont {Kodama}},\
  }\href@noop {} {\  (\bibinfo {year} {2015})},\ \Eprint
  {http://arxiv.org/abs/1506.03863} {arXiv:1506.03863 [nucl-th]} \BibitemShut
  {NoStop}%
\bibitem [{\citenamefont {Song}\ \emph {et~al.}(2017)\citenamefont {Song},
  \citenamefont {Zhou},\ and\ \citenamefont {Gajdosova}}]{Song:2017wtw}%
  \BibitemOpen
  \bibfield  {author} {\bibinfo {author} {\bibfnamefont {H.}~\bibnamefont
  {Song}}, \bibinfo {author} {\bibfnamefont {Y.}~\bibnamefont {Zhou}}, \ and\
  \bibinfo {author} {\bibfnamefont {K.}~\bibnamefont {Gajdosova}},\ }\href@noop
  {} {\  (\bibinfo {year} {2017})},\ \Eprint {http://arxiv.org/abs/1703.00670}
  {arXiv:1703.00670 [nucl-th]} \BibitemShut {NoStop}%
\bibitem [{\citenamefont {Adare}\ \emph {et~al.}(2013)\citenamefont {Adare}
  \emph {et~al.}}]{Adare:2013piz}%
  \BibitemOpen
  \bibfield  {author} {\bibinfo {author} {\bibfnamefont {A.}~\bibnamefont
  {Adare}} \emph {et~al.} (\bibinfo {collaboration} {PHENIX Collaboration}),\
  }\href@noop {} {\  (\bibinfo {year} {2013})},\ \Eprint
  {http://arxiv.org/abs/1303.1794} {arXiv:1303.1794 [nucl-ex]} \BibitemShut
  {NoStop}%
\bibitem [{\citenamefont {Chatrchyan}\ \emph
  {et~al.}(2013{\natexlab{a}})\citenamefont {Chatrchyan} \emph
  {et~al.}}]{CMS:2012qk}%
  \BibitemOpen
  \bibfield  {author} {\bibinfo {author} {\bibfnamefont {S.}~\bibnamefont
  {Chatrchyan}} \emph {et~al.} (\bibinfo {collaboration} {CMS Collaboration}),\
  }\href {\doibase 10.1016/j.physletb.2012.11.025} {\bibfield  {journal}
  {\bibinfo  {journal} {Phys.Lett.}\ }\textbf {\bibinfo {volume} {B718}},\
  \bibinfo {pages} {795} (\bibinfo {year} {2013}{\natexlab{a}})},\ \Eprint
  {http://arxiv.org/abs/1210.5482} {arXiv:1210.5482 [nucl-ex]} \BibitemShut
  {NoStop}%
\bibitem [{\citenamefont {Abelev}\ \emph
  {et~al.}(2013{\natexlab{a}})\citenamefont {Abelev} \emph
  {et~al.}}]{Abelev:2012ola}%
  \BibitemOpen
  \bibfield  {author} {\bibinfo {author} {\bibfnamefont {B.}~\bibnamefont
  {Abelev}} \emph {et~al.} (\bibinfo {collaboration} {ALICE Collaboration}),\
  }\href {\doibase 10.1016/j.physletb.2013.01.012} {\bibfield  {journal}
  {\bibinfo  {journal} {Phys.Lett.}\ }\textbf {\bibinfo {volume} {B719}},\
  \bibinfo {pages} {29} (\bibinfo {year} {2013}{\natexlab{a}})},\ \Eprint
  {http://arxiv.org/abs/1212.2001} {arXiv:1212.2001 [nucl-ex]} \BibitemShut
  {NoStop}%
\bibitem [{\citenamefont {Aad}\ \emph {et~al.}(2012)\citenamefont {Aad} \emph
  {et~al.}}]{Aad:2012gla}%
  \BibitemOpen
  \bibfield  {author} {\bibinfo {author} {\bibfnamefont {G.}~\bibnamefont
  {Aad}} \emph {et~al.} (\bibinfo {collaboration} {ATLAS Collaboration}),\
  }\href@noop {} {\  (\bibinfo {year} {2012})},\ \Eprint
  {http://arxiv.org/abs/1212.5198} {arXiv:1212.5198 [hep-ex]} \BibitemShut
  {NoStop}%
\bibitem [{\citenamefont {Dusling}\ \emph {et~al.}(2016)\citenamefont
  {Dusling}, \citenamefont {Li},\ and\ \citenamefont
  {Schenke}}]{Dusling:2015gta}%
  \BibitemOpen
  \bibfield  {author} {\bibinfo {author} {\bibfnamefont {K.}~\bibnamefont
  {Dusling}}, \bibinfo {author} {\bibfnamefont {W.}~\bibnamefont {Li}}, \ and\
  \bibinfo {author} {\bibfnamefont {B.}~\bibnamefont {Schenke}},\ }\href
  {\doibase 10.1142/S0218301316300022} {\bibfield  {journal} {\bibinfo
  {journal} {Int. J. Mod. Phys.}\ }\textbf {\bibinfo {volume} {E25}},\ \bibinfo
  {pages} {1630002} (\bibinfo {year} {2016})},\ \Eprint
  {http://arxiv.org/abs/1509.07939} {arXiv:1509.07939 [nucl-ex]} \BibitemShut
  {NoStop}%
\bibitem [{\citenamefont {Bozek}(2012)}]{Bozek:2011if}%
  \BibitemOpen
  \bibfield  {author} {\bibinfo {author} {\bibfnamefont {P.}~\bibnamefont
  {Bozek}},\ }\href {\doibase 10.1103/PhysRevC.85.014911} {\bibfield  {journal}
  {\bibinfo  {journal} {Phys.Rev.}\ }\textbf {\bibinfo {volume} {C85}},\
  \bibinfo {pages} {014911} (\bibinfo {year} {2012})},\ \Eprint
  {http://arxiv.org/abs/1112.0915} {arXiv:1112.0915 [hep-ph]} \BibitemShut
  {NoStop}%
\bibitem [{\citenamefont {Bozek}\ and\ \citenamefont
  {Broniowski}(2013{\natexlab{a}})}]{Bozek:2012gr}%
  \BibitemOpen
  \bibfield  {author} {\bibinfo {author} {\bibfnamefont {P.}~\bibnamefont
  {Bozek}}\ and\ \bibinfo {author} {\bibfnamefont {W.}~\bibnamefont
  {Broniowski}},\ }\href {\doibase 10.1016/j.physletb.2012.12.051} {\bibfield
  {journal} {\bibinfo  {journal} {Phys.Lett.}\ }\textbf {\bibinfo {volume}
  {B718}},\ \bibinfo {pages} {1557} (\bibinfo {year} {2013}{\natexlab{a}})},\
  \Eprint {http://arxiv.org/abs/1211.0845} {arXiv:1211.0845 [nucl-th]}
  \BibitemShut {NoStop}%
\bibitem [{\citenamefont {Bozek}\ and\ \citenamefont
  {Broniowski}(2013{\natexlab{b}})}]{Bozek:2013df}%
  \BibitemOpen
  \bibfield  {author} {\bibinfo {author} {\bibfnamefont {P.}~\bibnamefont
  {Bozek}}\ and\ \bibinfo {author} {\bibfnamefont {W.}~\bibnamefont
  {Broniowski}},\ }\href@noop {} {\  (\bibinfo {year} {2013}{\natexlab{b}})},\
  \Eprint {http://arxiv.org/abs/1301.3314} {arXiv:1301.3314 [nucl-th]}
  \BibitemShut {NoStop}%
\bibitem [{\citenamefont {Bozek}\ and\ \citenamefont
  {Broniowski}(2013{\natexlab{c}})}]{Bozek:2013uha}%
  \BibitemOpen
  \bibfield  {author} {\bibinfo {author} {\bibfnamefont {P.}~\bibnamefont
  {Bozek}}\ and\ \bibinfo {author} {\bibfnamefont {W.}~\bibnamefont
  {Broniowski}},\ }\href {\doibase 10.1103/PhysRevC.88.014903} {\bibfield
  {journal} {\bibinfo  {journal} {Phys.Rev.}\ }\textbf {\bibinfo {volume}
  {C88}},\ \bibinfo {pages} {014903} (\bibinfo {year} {2013}{\natexlab{c}})},\
  \Eprint {http://arxiv.org/abs/1304.3044} {arXiv:1304.3044 [nucl-th]}
  \BibitemShut {NoStop}%
\bibitem [{\citenamefont {Bozek}\ \emph {et~al.}(2013)\citenamefont {Bozek},
  \citenamefont {Broniowski},\ and\ \citenamefont {Torrieri}}]{Bozek:2013ska}%
  \BibitemOpen
  \bibfield  {author} {\bibinfo {author} {\bibfnamefont {P.}~\bibnamefont
  {Bozek}}, \bibinfo {author} {\bibfnamefont {W.}~\bibnamefont {Broniowski}}, \
  and\ \bibinfo {author} {\bibfnamefont {G.}~\bibnamefont {Torrieri}},\ }\href
  {\doibase 10.1103/PhysRevLett.111.172303} {\bibfield  {journal} {\bibinfo
  {journal} {Phys.Rev.Lett.}\ }\textbf {\bibinfo {volume} {111}},\ \bibinfo
  {pages} {172303} (\bibinfo {year} {2013})}\BibitemShut {NoStop}%
\bibitem [{\citenamefont {Qin}\ and\ \citenamefont
  {Muller}(2014)}]{Qin:2013bha}%
  \BibitemOpen
  \bibfield  {author} {\bibinfo {author} {\bibfnamefont {G.-Y.}\ \bibnamefont
  {Qin}}\ and\ \bibinfo {author} {\bibfnamefont {B.}~\bibnamefont {Muller}},\
  }\href {\doibase 10.1103/PhysRevC.89.044902} {\bibfield  {journal} {\bibinfo
  {journal} {Phys.Rev.}\ }\textbf {\bibinfo {volume} {C89}},\ \bibinfo {pages}
  {044902} (\bibinfo {year} {2014})},\ \Eprint {http://arxiv.org/abs/1306.3439}
  {arXiv:1306.3439 [nucl-th]} \BibitemShut {NoStop}%
\bibitem [{\citenamefont {Werner}\ \emph {et~al.}(2014)\citenamefont {Werner},
  \citenamefont {Bleicher}, \citenamefont {Guiot}, \citenamefont {Karpenko},\
  and\ \citenamefont {Pierog}}]{Werner:2013ipa}%
  \BibitemOpen
  \bibfield  {author} {\bibinfo {author} {\bibfnamefont {K.}~\bibnamefont
  {Werner}}, \bibinfo {author} {\bibfnamefont {M.}~\bibnamefont {Bleicher}},
  \bibinfo {author} {\bibfnamefont {B.}~\bibnamefont {Guiot}}, \bibinfo
  {author} {\bibfnamefont {I.}~\bibnamefont {Karpenko}}, \ and\ \bibinfo
  {author} {\bibfnamefont {T.}~\bibnamefont {Pierog}},\ }\href {\doibase
  10.1103/PhysRevLett.112.232301} {\bibfield  {journal} {\bibinfo  {journal}
  {Phys. Rev. Lett.}\ }\textbf {\bibinfo {volume} {112}},\ \bibinfo {pages}
  {232301} (\bibinfo {year} {2014})},\ \Eprint {http://arxiv.org/abs/1307.4379}
  {arXiv:1307.4379 [nucl-th]} \BibitemShut {NoStop}%
\bibitem [{\citenamefont {Kozlov}\ \emph {et~al.}(2014)\citenamefont {Kozlov},
  \citenamefont {Luzum}, \citenamefont {Denicol}, \citenamefont {Jeon},\ and\
  \citenamefont {Gale}}]{Kozlov:2014fqa}%
  \BibitemOpen
  \bibfield  {author} {\bibinfo {author} {\bibfnamefont {I.}~\bibnamefont
  {Kozlov}}, \bibinfo {author} {\bibfnamefont {M.}~\bibnamefont {Luzum}},
  \bibinfo {author} {\bibfnamefont {G.}~\bibnamefont {Denicol}}, \bibinfo
  {author} {\bibfnamefont {S.}~\bibnamefont {Jeon}}, \ and\ \bibinfo {author}
  {\bibfnamefont {C.}~\bibnamefont {Gale}},\ }\href@noop {} {\  (\bibinfo
  {year} {2014})},\ \Eprint {http://arxiv.org/abs/1405.3976} {arXiv:1405.3976
  [nucl-th]} \BibitemShut {NoStop}%
\bibitem [{\citenamefont {Romatschke}(2015)}]{Romatschke:2015gxa}%
  \BibitemOpen
  \bibfield  {author} {\bibinfo {author} {\bibfnamefont {P.}~\bibnamefont
  {Romatschke}},\ }\href {\doibase 10.1140/epjc/s10052-015-3509-3} {\bibfield
  {journal} {\bibinfo  {journal} {Eur. Phys. J.}\ }\textbf {\bibinfo {volume}
  {C75}},\ \bibinfo {pages} {305} (\bibinfo {year} {2015})},\ \Eprint
  {http://arxiv.org/abs/1502.04745} {arXiv:1502.04745 [nucl-th]} \BibitemShut
  {NoStop}%
\bibitem [{\citenamefont {Shen}\ \emph {et~al.}(2017)\citenamefont {Shen},
  \citenamefont {Paquet}, \citenamefont {Denicol}, \citenamefont {Jeon},\ and\
  \citenamefont {Gale}}]{Shen:2016zpp}%
  \BibitemOpen
  \bibfield  {author} {\bibinfo {author} {\bibfnamefont {C.}~\bibnamefont
  {Shen}}, \bibinfo {author} {\bibfnamefont {J.-F.}\ \bibnamefont {Paquet}},
  \bibinfo {author} {\bibfnamefont {G.~S.}\ \bibnamefont {Denicol}}, \bibinfo
  {author} {\bibfnamefont {S.}~\bibnamefont {Jeon}}, \ and\ \bibinfo {author}
  {\bibfnamefont {C.}~\bibnamefont {Gale}},\ }\href {\doibase
  10.1103/PhysRevC.95.014906} {\bibfield  {journal} {\bibinfo  {journal} {Phys.
  Rev.}\ }\textbf {\bibinfo {volume} {C95}},\ \bibinfo {pages} {014906}
  (\bibinfo {year} {2017})},\ \Eprint {http://arxiv.org/abs/1609.02590}
  {arXiv:1609.02590 [nucl-th]} \BibitemShut {NoStop}%
\bibitem [{\citenamefont {Weller}\ and\ \citenamefont
  {Romatschke}(2017)}]{Weller:2017tsr}%
  \BibitemOpen
  \bibfield  {author} {\bibinfo {author} {\bibfnamefont {R.~D.}\ \bibnamefont
  {Weller}}\ and\ \bibinfo {author} {\bibfnamefont {P.}~\bibnamefont
  {Romatschke}},\ }\href@noop {} {\  (\bibinfo {year} {2017})},\ \Eprint
  {http://arxiv.org/abs/1701.07145} {arXiv:1701.07145 [nucl-th]} \BibitemShut
  {NoStop}%
\bibitem [{\citenamefont {Schlichting}\ and\ \citenamefont
  {Tribedy}(2016)}]{Schlichting:2016kjw}%
  \BibitemOpen
  \bibfield  {author} {\bibinfo {author} {\bibfnamefont {S.}~\bibnamefont
  {Schlichting}}\ and\ \bibinfo {author} {\bibfnamefont {P.}~\bibnamefont
  {Tribedy}},\ }\href {\doibase 10.1155/2016/8460349} {\bibfield  {journal}
  {\bibinfo  {journal} {Adv. High Energy Phys.}\ }\textbf {\bibinfo {volume}
  {2016}},\ \bibinfo {pages} {8460349} (\bibinfo {year} {2016})}\BibitemShut
  {NoStop}%
\bibitem [{\citenamefont {Schenke}\ \emph {et~al.}(2016)\citenamefont
  {Schenke}, \citenamefont {Schlichting}, \citenamefont {Tribedy},\ and\
  \citenamefont {Venugopalan}}]{Schenke:2016lrs}%
  \BibitemOpen
  \bibfield  {author} {\bibinfo {author} {\bibfnamefont {B.}~\bibnamefont
  {Schenke}}, \bibinfo {author} {\bibfnamefont {S.}~\bibnamefont
  {Schlichting}}, \bibinfo {author} {\bibfnamefont {P.}~\bibnamefont
  {Tribedy}}, \ and\ \bibinfo {author} {\bibfnamefont {R.}~\bibnamefont
  {Venugopalan}},\ }\href {\doibase 10.1103/PhysRevLett.117.162301} {\bibfield
  {journal} {\bibinfo  {journal} {Phys. Rev. Lett.}\ }\textbf {\bibinfo
  {volume} {117}},\ \bibinfo {pages} {162301} (\bibinfo {year} {2016})},\
  \Eprint {http://arxiv.org/abs/1607.02496} {arXiv:1607.02496 [hep-ph]}
  \BibitemShut {NoStop}%
\bibitem [{\citenamefont {Dusling}\ \emph {et~al.}(2017)\citenamefont
  {Dusling}, \citenamefont {Mace},\ and\ \citenamefont
  {Venugopalan}}]{Dusling:2017dqg}%
  \BibitemOpen
  \bibfield  {author} {\bibinfo {author} {\bibfnamefont {K.}~\bibnamefont
  {Dusling}}, \bibinfo {author} {\bibfnamefont {M.}~\bibnamefont {Mace}}, \
  and\ \bibinfo {author} {\bibfnamefont {R.}~\bibnamefont {Venugopalan}},\
  }\href@noop {} {\  (\bibinfo {year} {2017})},\ \Eprint
  {http://arxiv.org/abs/1705.00745} {arXiv:1705.00745 [hep-ph]} \BibitemShut
  {NoStop}%
\bibitem [{\citenamefont {Schenke}\ \emph {et~al.}(2015)\citenamefont
  {Schenke}, \citenamefont {Schlichting},\ and\ \citenamefont
  {Venugopalan}}]{Schenke:2015aqa}%
  \BibitemOpen
  \bibfield  {author} {\bibinfo {author} {\bibfnamefont {B.}~\bibnamefont
  {Schenke}}, \bibinfo {author} {\bibfnamefont {S.}~\bibnamefont
  {Schlichting}}, \ and\ \bibinfo {author} {\bibfnamefont {R.}~\bibnamefont
  {Venugopalan}},\ }\href {\doibase 10.1016/j.physletb.2015.05.051} {\bibfield
  {journal} {\bibinfo  {journal} {Phys. Lett.}\ }\textbf {\bibinfo {volume}
  {B747}},\ \bibinfo {pages} {76} (\bibinfo {year} {2015})},\ \Eprint
  {http://arxiv.org/abs/1502.01331} {arXiv:1502.01331 [hep-ph]} \BibitemShut
  {NoStop}%
\bibitem [{\citenamefont {McLerran}\ and\ \citenamefont
  {Skokov}(2017)}]{McLerran:2016snu}%
  \BibitemOpen
  \bibfield  {author} {\bibinfo {author} {\bibfnamefont {L.}~\bibnamefont
  {McLerran}}\ and\ \bibinfo {author} {\bibfnamefont {V.}~\bibnamefont
  {Skokov}},\ }\href {\doibase 10.1016/j.nuclphysa.2016.12.011} {\bibfield
  {journal} {\bibinfo  {journal} {Nucl. Phys.}\ }\textbf {\bibinfo {volume}
  {A959}},\ \bibinfo {pages} {83} (\bibinfo {year} {2017})},\ \Eprint
  {http://arxiv.org/abs/1611.09870} {arXiv:1611.09870 [hep-ph]} \BibitemShut
  {NoStop}%
\bibitem [{\citenamefont {Chesler}\ and\ \citenamefont
  {Yaffe}(2010)}]{Chesler:2009cy}%
  \BibitemOpen
  \bibfield  {author} {\bibinfo {author} {\bibfnamefont {P.~M.}\ \bibnamefont
  {Chesler}}\ and\ \bibinfo {author} {\bibfnamefont {L.~G.}\ \bibnamefont
  {Yaffe}},\ }\href {\doibase 10.1103/PhysRevD.82.026006} {\bibfield  {journal}
  {\bibinfo  {journal} {Phys. Rev.}\ }\textbf {\bibinfo {volume} {D82}},\
  \bibinfo {pages} {026006} (\bibinfo {year} {2010})},\ \Eprint
  {http://arxiv.org/abs/0906.4426} {arXiv:0906.4426 [hep-th]} \BibitemShut
  {NoStop}%
\bibitem [{\citenamefont {Wu}\ and\ \citenamefont
  {Romatschke}(2011)}]{Wu:2011yd}%
  \BibitemOpen
  \bibfield  {author} {\bibinfo {author} {\bibfnamefont {B.}~\bibnamefont
  {Wu}}\ and\ \bibinfo {author} {\bibfnamefont {P.}~\bibnamefont
  {Romatschke}},\ }\href {\doibase 10.1142/S0129183111016920} {\bibfield
  {journal} {\bibinfo  {journal} {Int. J. Mod. Phys.}\ }\textbf {\bibinfo
  {volume} {C22}},\ \bibinfo {pages} {1317} (\bibinfo {year} {2011})},\ \Eprint
  {http://arxiv.org/abs/1108.3715} {arXiv:1108.3715 [hep-th]} \BibitemShut
  {NoStop}%
\bibitem [{\citenamefont {Heller}\ \emph {et~al.}(2012)\citenamefont {Heller},
  \citenamefont {Janik},\ and\ \citenamefont {Witaszczyk}}]{Heller:2011ju}%
  \BibitemOpen
  \bibfield  {author} {\bibinfo {author} {\bibfnamefont {M.~P.}\ \bibnamefont
  {Heller}}, \bibinfo {author} {\bibfnamefont {R.~A.}\ \bibnamefont {Janik}}, \
  and\ \bibinfo {author} {\bibfnamefont {P.}~\bibnamefont {Witaszczyk}},\
  }\href {\doibase 10.1103/PhysRevLett.108.201602} {\bibfield  {journal}
  {\bibinfo  {journal} {Phys. Rev. Lett.}\ }\textbf {\bibinfo {volume} {108}},\
  \bibinfo {pages} {201602} (\bibinfo {year} {2012})},\ \Eprint
  {http://arxiv.org/abs/1103.3452} {arXiv:1103.3452 [hep-th]} \BibitemShut
  {NoStop}%
\bibitem [{\citenamefont {van~der Schee}(2013)}]{vanderSchee:2012qj}%
  \BibitemOpen
  \bibfield  {author} {\bibinfo {author} {\bibfnamefont {W.}~\bibnamefont
  {van~der Schee}},\ }\href {\doibase 10.1103/PhysRevD.87.061901} {\bibfield
  {journal} {\bibinfo  {journal} {Phys. Rev.}\ }\textbf {\bibinfo {volume}
  {D87}},\ \bibinfo {pages} {061901} (\bibinfo {year} {2013})},\ \Eprint
  {http://arxiv.org/abs/1211.2218} {arXiv:1211.2218 [hep-th]} \BibitemShut
  {NoStop}%
\bibitem [{\citenamefont {Casalderrey-Solana}\ \emph
  {et~al.}(2013)\citenamefont {Casalderrey-Solana}, \citenamefont {Heller},
  \citenamefont {Mateos},\ and\ \citenamefont {van~der
  Schee}}]{Casalderrey-Solana:2013aba}%
  \BibitemOpen
  \bibfield  {author} {\bibinfo {author} {\bibfnamefont {J.}~\bibnamefont
  {Casalderrey-Solana}}, \bibinfo {author} {\bibfnamefont {M.~P.}\ \bibnamefont
  {Heller}}, \bibinfo {author} {\bibfnamefont {D.}~\bibnamefont {Mateos}}, \
  and\ \bibinfo {author} {\bibfnamefont {W.}~\bibnamefont {van~der Schee}},\
  }\href {\doibase 10.1103/PhysRevLett.111.181601} {\bibfield  {journal}
  {\bibinfo  {journal} {Phys. Rev. Lett.}\ }\textbf {\bibinfo {volume} {111}},\
  \bibinfo {pages} {181601} (\bibinfo {year} {2013})},\ \Eprint
  {http://arxiv.org/abs/1305.4919} {arXiv:1305.4919 [hep-th]} \BibitemShut
  {NoStop}%
\bibitem [{\citenamefont {Kurkela}\ and\ \citenamefont
  {Zhu}(2015)}]{Kurkela:2015qoa}%
  \BibitemOpen
  \bibfield  {author} {\bibinfo {author} {\bibfnamefont {A.}~\bibnamefont
  {Kurkela}}\ and\ \bibinfo {author} {\bibfnamefont {Y.}~\bibnamefont {Zhu}},\
  }\href {\doibase 10.1103/PhysRevLett.115.182301} {\bibfield  {journal}
  {\bibinfo  {journal} {Phys. Rev. Lett.}\ }\textbf {\bibinfo {volume} {115}},\
  \bibinfo {pages} {182301} (\bibinfo {year} {2015})},\ \Eprint
  {http://arxiv.org/abs/1506.06647} {arXiv:1506.06647 [hep-ph]} \BibitemShut
  {NoStop}%
\bibitem [{\citenamefont {Keegan}\ \emph
  {et~al.}(2016{\natexlab{a}})\citenamefont {Keegan}, \citenamefont {Kurkela},
  \citenamefont {Romatschke}, \citenamefont {van~der Schee},\ and\
  \citenamefont {Zhu}}]{Keegan:2015avk}%
  \BibitemOpen
  \bibfield  {author} {\bibinfo {author} {\bibfnamefont {L.}~\bibnamefont
  {Keegan}}, \bibinfo {author} {\bibfnamefont {A.}~\bibnamefont {Kurkela}},
  \bibinfo {author} {\bibfnamefont {P.}~\bibnamefont {Romatschke}}, \bibinfo
  {author} {\bibfnamefont {W.}~\bibnamefont {van~der Schee}}, \ and\ \bibinfo
  {author} {\bibfnamefont {Y.}~\bibnamefont {Zhu}},\ }\href {\doibase
  10.1007/JHEP04(2016)031} {\bibfield  {journal} {\bibinfo  {journal} {JHEP}\
  }\textbf {\bibinfo {volume} {04}},\ \bibinfo {pages} {031} (\bibinfo {year}
  {2016}{\natexlab{a}})},\ \Eprint {http://arxiv.org/abs/1512.05347}
  {arXiv:1512.05347 [hep-th]} \BibitemShut {NoStop}%
\bibitem [{\citenamefont {Chesler}(2016)}]{Chesler:2016ceu}%
  \BibitemOpen
  \bibfield  {author} {\bibinfo {author} {\bibfnamefont {P.~M.}\ \bibnamefont
  {Chesler}},\ }\href {\doibase 10.1007/JHEP03(2016)146} {\bibfield  {journal}
  {\bibinfo  {journal} {JHEP}\ }\textbf {\bibinfo {volume} {03}},\ \bibinfo
  {pages} {146} (\bibinfo {year} {2016})},\ \Eprint
  {http://arxiv.org/abs/1601.01583} {arXiv:1601.01583 [hep-th]} \BibitemShut
  {NoStop}%
\bibitem [{\citenamefont {Attems}\ \emph {et~al.}(2017)\citenamefont {Attems},
  \citenamefont {Casalderrey-Solana}, \citenamefont {Mateos}, \citenamefont
  {Santos-Olivan}, \citenamefont {Sopuerta}, \citenamefont {Triana},\ and\
  \citenamefont {Zilhao}}]{Attems:2016tby}%
  \BibitemOpen
  \bibfield  {author} {\bibinfo {author} {\bibfnamefont {M.}~\bibnamefont
  {Attems}}, \bibinfo {author} {\bibfnamefont {J.}~\bibnamefont
  {Casalderrey-Solana}}, \bibinfo {author} {\bibfnamefont {D.}~\bibnamefont
  {Mateos}}, \bibinfo {author} {\bibfnamefont {D.}~\bibnamefont
  {Santos-Olivan}}, \bibinfo {author} {\bibfnamefont {C.~F.}\ \bibnamefont
  {Sopuerta}}, \bibinfo {author} {\bibfnamefont {M.}~\bibnamefont {Triana}}, \
  and\ \bibinfo {author} {\bibfnamefont {M.}~\bibnamefont {Zilhao}},\ }\href
  {\doibase 10.1007/JHEP01(2017)026} {\bibfield  {journal} {\bibinfo  {journal}
  {JHEP}\ }\textbf {\bibinfo {volume} {01}},\ \bibinfo {pages} {026} (\bibinfo
  {year} {2017})},\ \Eprint {http://arxiv.org/abs/1604.06439} {arXiv:1604.06439
  [hep-th]} \BibitemShut {NoStop}%
\bibitem [{\citenamefont
  {Romatschke}(2017{\natexlab{a}})}]{Romatschke:2016hle}%
  \BibitemOpen
  \bibfield  {author} {\bibinfo {author} {\bibfnamefont {P.}~\bibnamefont
  {Romatschke}},\ }\href {\doibase 10.1140/epjc/s10052-016-4567-x} {\bibfield
  {journal} {\bibinfo  {journal} {Eur. Phys. J.}\ }\textbf {\bibinfo {volume}
  {C77}},\ \bibinfo {pages} {21} (\bibinfo {year} {2017}{\natexlab{a}})},\
  \Eprint {http://arxiv.org/abs/1609.02820} {arXiv:1609.02820 [nucl-th]}
  \BibitemShut {NoStop}%
\bibitem [{\citenamefont
  {Romatschke}(2017{\natexlab{b}})}]{Romatschke:2017vte}%
  \BibitemOpen
  \bibfield  {author} {\bibinfo {author} {\bibfnamefont {P.}~\bibnamefont
  {Romatschke}},\ }\href@noop {} {\  (\bibinfo {year} {2017}{\natexlab{b}})},\
  \Eprint {http://arxiv.org/abs/1704.08699} {arXiv:1704.08699 [hep-th]}
  \BibitemShut {NoStop}%
\bibitem [{\citenamefont {Schenke}\ \emph
  {et~al.}(2012{\natexlab{a}})\citenamefont {Schenke}, \citenamefont
  {Tribedy},\ and\ \citenamefont {Venugopalan}}]{Schenke:2012wb}%
  \BibitemOpen
  \bibfield  {author} {\bibinfo {author} {\bibfnamefont {B.}~\bibnamefont
  {Schenke}}, \bibinfo {author} {\bibfnamefont {P.}~\bibnamefont {Tribedy}}, \
  and\ \bibinfo {author} {\bibfnamefont {R.}~\bibnamefont {Venugopalan}},\
  }\href {\doibase 10.1103/PhysRevLett.108.252301} {\bibfield  {journal}
  {\bibinfo  {journal} {Phys. Rev. Lett.}\ }\textbf {\bibinfo {volume} {108}},\
  \bibinfo {pages} {252301} (\bibinfo {year} {2012}{\natexlab{a}})},\ \Eprint
  {http://arxiv.org/abs/1202.6646} {arXiv:1202.6646 [nucl-th]} \BibitemShut
  {NoStop}%
\bibitem [{\citenamefont {Schenke}\ \emph
  {et~al.}(2012{\natexlab{b}})\citenamefont {Schenke}, \citenamefont
  {Tribedy},\ and\ \citenamefont {Venugopalan}}]{Schenke:2012hg}%
  \BibitemOpen
  \bibfield  {author} {\bibinfo {author} {\bibfnamefont {B.}~\bibnamefont
  {Schenke}}, \bibinfo {author} {\bibfnamefont {P.}~\bibnamefont {Tribedy}}, \
  and\ \bibinfo {author} {\bibfnamefont {R.}~\bibnamefont {Venugopalan}},\
  }\href {\doibase 10.1103/PhysRevC.86.034908} {\bibfield  {journal} {\bibinfo
  {journal} {Phys. Rev.}\ }\textbf {\bibinfo {volume} {C86}},\ \bibinfo {pages}
  {034908} (\bibinfo {year} {2012}{\natexlab{b}})},\ \Eprint
  {http://arxiv.org/abs/1206.6805} {arXiv:1206.6805 [hep-ph]} \BibitemShut
  {NoStop}%
\bibitem [{\citenamefont {Gale}\ \emph
  {et~al.}(2013{\natexlab{b}})\citenamefont {Gale}, \citenamefont {Jeon},
  \citenamefont {Schenke}, \citenamefont {Tribedy},\ and\ \citenamefont
  {Venugopalan}}]{Gale:2012rq}%
  \BibitemOpen
  \bibfield  {author} {\bibinfo {author} {\bibfnamefont {C.}~\bibnamefont
  {Gale}}, \bibinfo {author} {\bibfnamefont {S.}~\bibnamefont {Jeon}}, \bibinfo
  {author} {\bibfnamefont {B.}~\bibnamefont {Schenke}}, \bibinfo {author}
  {\bibfnamefont {P.}~\bibnamefont {Tribedy}}, \ and\ \bibinfo {author}
  {\bibfnamefont {R.}~\bibnamefont {Venugopalan}},\ }\href {\doibase
  10.1103/PhysRevLett.110.012302} {\bibfield  {journal} {\bibinfo  {journal}
  {Phys. Rev. Lett.}\ }\textbf {\bibinfo {volume} {110}},\ \bibinfo {pages}
  {012302} (\bibinfo {year} {2013}{\natexlab{b}})},\ \Eprint
  {http://arxiv.org/abs/1209.6330} {arXiv:1209.6330 [nucl-th]} \BibitemShut
  {NoStop}%
\bibitem [{\citenamefont {Schenke}\ and\ \citenamefont
  {Venugopalan}(2014)}]{Schenke:2014zha}%
  \BibitemOpen
  \bibfield  {author} {\bibinfo {author} {\bibfnamefont {B.}~\bibnamefont
  {Schenke}}\ and\ \bibinfo {author} {\bibfnamefont {R.}~\bibnamefont
  {Venugopalan}},\ }\href {\doibase 10.1103/PhysRevLett.113.102301} {\bibfield
  {journal} {\bibinfo  {journal} {Phys.Rev.Lett.}\ }\textbf {\bibinfo {volume}
  {113}},\ \bibinfo {pages} {102301} (\bibinfo {year} {2014})},\ \Eprint
  {http://arxiv.org/abs/1405.3605} {arXiv:1405.3605 [nucl-th]} \BibitemShut
  {NoStop}%
\bibitem [{\citenamefont {Bernhard}\ \emph {et~al.}(2016)\citenamefont
  {Bernhard}, \citenamefont {Moreland}, \citenamefont {Bass}, \citenamefont
  {Liu},\ and\ \citenamefont {Heinz}}]{Bernhard:2016tnd}%
  \BibitemOpen
  \bibfield  {author} {\bibinfo {author} {\bibfnamefont {J.~E.}\ \bibnamefont
  {Bernhard}}, \bibinfo {author} {\bibfnamefont {J.~S.}\ \bibnamefont
  {Moreland}}, \bibinfo {author} {\bibfnamefont {S.~A.}\ \bibnamefont {Bass}},
  \bibinfo {author} {\bibfnamefont {J.}~\bibnamefont {Liu}}, \ and\ \bibinfo
  {author} {\bibfnamefont {U.}~\bibnamefont {Heinz}},\ }\href {\doibase
  10.1103/PhysRevC.94.024907} {\bibfield  {journal} {\bibinfo  {journal} {Phys.
  Rev.}\ }\textbf {\bibinfo {volume} {C94}},\ \bibinfo {pages} {024907}
  (\bibinfo {year} {2016})},\ \Eprint {http://arxiv.org/abs/1605.03954}
  {arXiv:1605.03954 [nucl-th]} \BibitemShut {NoStop}%
\bibitem [{\citenamefont {Niemi}\ \emph {et~al.}(2013)\citenamefont {Niemi},
  \citenamefont {Denicol}, \citenamefont {Holopainen},\ and\ \citenamefont
  {Huovinen}}]{Niemi:2012aj}%
  \BibitemOpen
  \bibfield  {author} {\bibinfo {author} {\bibfnamefont {H.}~\bibnamefont
  {Niemi}}, \bibinfo {author} {\bibfnamefont {G.~S.}\ \bibnamefont {Denicol}},
  \bibinfo {author} {\bibfnamefont {H.}~\bibnamefont {Holopainen}}, \ and\
  \bibinfo {author} {\bibfnamefont {P.}~\bibnamefont {Huovinen}},\ }\href
  {\doibase 10.1103/PhysRevC.87.054901} {\bibfield  {journal} {\bibinfo
  {journal} {Phys. Rev.}\ }\textbf {\bibinfo {volume} {C87}},\ \bibinfo {pages}
  {054901} (\bibinfo {year} {2013})},\ \Eprint {http://arxiv.org/abs/1212.1008}
  {arXiv:1212.1008 [nucl-th]} \BibitemShut {NoStop}%
\bibitem [{\citenamefont {Aad}\ \emph {et~al.}(2013)\citenamefont {Aad} \emph
  {et~al.}}]{Aad:2013xma}%
  \BibitemOpen
  \bibfield  {author} {\bibinfo {author} {\bibfnamefont {G.}~\bibnamefont
  {Aad}} \emph {et~al.} (\bibinfo {collaboration} {ATLAS}),\ }\href {\doibase
  10.1007/JHEP11(2013)183} {\bibfield  {journal} {\bibinfo  {journal} {JHEP}\
  }\textbf {\bibinfo {volume} {11}},\ \bibinfo {pages} {183} (\bibinfo {year}
  {2013})},\ \Eprint {http://arxiv.org/abs/1305.2942} {arXiv:1305.2942
  [hep-ex]} \BibitemShut {NoStop}%
\bibitem [{\citenamefont {Niemi}\ \emph {et~al.}(2015)\citenamefont {Niemi},
  \citenamefont {Eskola},\ and\ \citenamefont {Paatelainen}}]{Niemi:2015qia}%
  \BibitemOpen
  \bibfield  {author} {\bibinfo {author} {\bibfnamefont {H.}~\bibnamefont
  {Niemi}}, \bibinfo {author} {\bibfnamefont {K.~J.}\ \bibnamefont {Eskola}}, \
  and\ \bibinfo {author} {\bibfnamefont {R.}~\bibnamefont {Paatelainen}},\
  }\href@noop {} {\  (\bibinfo {year} {2015})},\ \Eprint
  {http://arxiv.org/abs/1505.02677} {arXiv:1505.02677 [hep-ph]} \BibitemShut
  {NoStop}%
\bibitem [{\citenamefont {Moreland}\ \emph {et~al.}(2015)\citenamefont
  {Moreland}, \citenamefont {Bernhard},\ and\ \citenamefont
  {Bass}}]{Moreland:2014oya}%
  \BibitemOpen
  \bibfield  {author} {\bibinfo {author} {\bibfnamefont {J.~S.}\ \bibnamefont
  {Moreland}}, \bibinfo {author} {\bibfnamefont {J.~E.}\ \bibnamefont
  {Bernhard}}, \ and\ \bibinfo {author} {\bibfnamefont {S.~A.}\ \bibnamefont
  {Bass}},\ }\href {\doibase 10.1103/PhysRevC.92.011901} {\bibfield  {journal}
  {\bibinfo  {journal} {Phys. Rev.}\ }\textbf {\bibinfo {volume} {C92}},\
  \bibinfo {pages} {011901} (\bibinfo {year} {2015})},\ \Eprint
  {http://arxiv.org/abs/1412.4708} {arXiv:1412.4708 [nucl-th]} \BibitemShut
  {NoStop}%
\bibitem [{\citenamefont {Schenke}\ \emph {et~al.}(2010)\citenamefont
  {Schenke}, \citenamefont {Jeon},\ and\ \citenamefont
  {Gale}}]{Schenke:2010nt}%
  \BibitemOpen
  \bibfield  {author} {\bibinfo {author} {\bibfnamefont {B.}~\bibnamefont
  {Schenke}}, \bibinfo {author} {\bibfnamefont {S.}~\bibnamefont {Jeon}}, \
  and\ \bibinfo {author} {\bibfnamefont {C.}~\bibnamefont {Gale}},\ }\href
  {\doibase 10.1103/PhysRevC.82.014903} {\bibfield  {journal} {\bibinfo
  {journal} {Phys. Rev.}\ }\textbf {\bibinfo {volume} {C82}},\ \bibinfo {pages}
  {014903} (\bibinfo {year} {2010})},\ \Eprint {http://arxiv.org/abs/1004.1408}
  {arXiv:1004.1408 [hep-ph]} \BibitemShut {NoStop}%
\bibitem [{\citenamefont {Schenke}\ \emph {et~al.}(2011)\citenamefont
  {Schenke}, \citenamefont {Jeon},\ and\ \citenamefont
  {Gale}}]{Schenke:2010rr}%
  \BibitemOpen
  \bibfield  {author} {\bibinfo {author} {\bibfnamefont {B.}~\bibnamefont
  {Schenke}}, \bibinfo {author} {\bibfnamefont {S.}~\bibnamefont {Jeon}}, \
  and\ \bibinfo {author} {\bibfnamefont {C.}~\bibnamefont {Gale}},\ }\href
  {\doibase 10.1103/PhysRevLett.106.042301} {\bibfield  {journal} {\bibinfo
  {journal} {Phys. Rev. Lett.}\ }\textbf {\bibinfo {volume} {106}},\ \bibinfo
  {pages} {042301} (\bibinfo {year} {2011})},\ \Eprint
  {http://arxiv.org/abs/1009.3244} {arXiv:1009.3244 [hep-ph]} \BibitemShut
  {NoStop}%
\bibitem [{\citenamefont {Schenke}\ \emph
  {et~al.}(2012{\natexlab{c}})\citenamefont {Schenke}, \citenamefont {Jeon},\
  and\ \citenamefont {Gale}}]{Schenke:2011bn}%
  \BibitemOpen
  \bibfield  {author} {\bibinfo {author} {\bibfnamefont {B.}~\bibnamefont
  {Schenke}}, \bibinfo {author} {\bibfnamefont {S.}~\bibnamefont {Jeon}}, \
  and\ \bibinfo {author} {\bibfnamefont {C.}~\bibnamefont {Gale}},\ }\href
  {\doibase 10.1103/PhysRevC.85.024901} {\bibfield  {journal} {\bibinfo
  {journal} {Phys. Rev.}\ }\textbf {\bibinfo {volume} {C85}},\ \bibinfo {pages}
  {024901} (\bibinfo {year} {2012}{\natexlab{c}})},\ \Eprint
  {http://arxiv.org/abs/1109.6289} {arXiv:1109.6289 [hep-ph]} \BibitemShut
  {NoStop}%
\bibitem [{\citenamefont {Bass}\ \emph {et~al.}(1998)\citenamefont {Bass} \emph
  {et~al.}}]{Bass:1998ca}%
  \BibitemOpen
  \bibfield  {author} {\bibinfo {author} {\bibfnamefont {S.~A.}\ \bibnamefont
  {Bass}} \emph {et~al.},\ }\href {\doibase 10.1016/S0146-6410(98)00058-1}
  {\bibfield  {journal} {\bibinfo  {journal} {Prog. Part. Nucl. Phys.}\
  }\textbf {\bibinfo {volume} {41}},\ \bibinfo {pages} {255} (\bibinfo {year}
  {1998})},\ \bibinfo {note} {[Prog. Part. Nucl. Phys.41,225(1998)]},\ \Eprint
  {http://arxiv.org/abs/nucl-th/9803035} {arXiv:nucl-th/9803035 [nucl-th]}
  \BibitemShut {NoStop}%
\bibitem [{\citenamefont {Bleicher}\ \emph {et~al.}(1999)\citenamefont
  {Bleicher} \emph {et~al.}}]{Bleicher:1999xi}%
  \BibitemOpen
  \bibfield  {author} {\bibinfo {author} {\bibfnamefont {M.}~\bibnamefont
  {Bleicher}} \emph {et~al.},\ }\href {\doibase 10.1088/0954-3899/25/9/308}
  {\bibfield  {journal} {\bibinfo  {journal} {J. Phys.}\ }\textbf {\bibinfo
  {volume} {G25}},\ \bibinfo {pages} {1859} (\bibinfo {year} {1999})},\ \Eprint
  {http://arxiv.org/abs/hep-ph/9909407} {arXiv:hep-ph/9909407 [hep-ph]}
  \BibitemShut {NoStop}%
\bibitem [{\citenamefont {Bzdak}\ \emph {et~al.}(2013)\citenamefont {Bzdak},
  \citenamefont {Schenke}, \citenamefont {Tribedy},\ and\ \citenamefont
  {Venugopalan}}]{Bzdak:2013zma}%
  \BibitemOpen
  \bibfield  {author} {\bibinfo {author} {\bibfnamefont {A.}~\bibnamefont
  {Bzdak}}, \bibinfo {author} {\bibfnamefont {B.}~\bibnamefont {Schenke}},
  \bibinfo {author} {\bibfnamefont {P.}~\bibnamefont {Tribedy}}, \ and\
  \bibinfo {author} {\bibfnamefont {R.}~\bibnamefont {Venugopalan}},\
  }\href@noop {} {\  (\bibinfo {year} {2013})},\ \Eprint
  {http://arxiv.org/abs/1304.3403} {arXiv:1304.3403 [nucl-th]} \BibitemShut
  {NoStop}%
\bibitem [{\citenamefont {Rezaeian}\ \emph {et~al.}(2013)\citenamefont
  {Rezaeian}, \citenamefont {Siddikov}, \citenamefont {Van~de Klundert},\ and\
  \citenamefont {Venugopalan}}]{Rezaeian:2012ji}%
  \BibitemOpen
  \bibfield  {author} {\bibinfo {author} {\bibfnamefont {A.~H.}\ \bibnamefont
  {Rezaeian}}, \bibinfo {author} {\bibfnamefont {M.}~\bibnamefont {Siddikov}},
  \bibinfo {author} {\bibfnamefont {M.}~\bibnamefont {Van~de Klundert}}, \ and\
  \bibinfo {author} {\bibfnamefont {R.}~\bibnamefont {Venugopalan}},\
  }\href@noop {} {\bibfield  {journal} {\bibinfo  {journal} {Phys.Rev.}\
  }\textbf {\bibinfo {volume} {D87}},\ \bibinfo {pages} {034002} (\bibinfo
  {year} {2013})},\ \Eprint {http://arxiv.org/abs/1212.2974} {arXiv:1212.2974
  [hep-ph]} \BibitemShut {NoStop}%
\bibitem [{\citenamefont {Breitweg}\ \emph {et~al.}(2000)\citenamefont
  {Breitweg} \emph {et~al.}}]{Breitweg:1999jy}%
  \BibitemOpen
  \bibfield  {author} {\bibinfo {author} {\bibfnamefont {J.}~\bibnamefont
  {Breitweg}} \emph {et~al.} (\bibinfo {collaboration} {ZEUS}),\ }\href
  {\doibase 10.1007/s100520050748} {\bibfield  {journal} {\bibinfo  {journal}
  {Eur. Phys. J.}\ }\textbf {\bibinfo {volume} {C14}},\ \bibinfo {pages} {213}
  (\bibinfo {year} {2000})},\ \Eprint {http://arxiv.org/abs/hep-ex/9910038}
  {arXiv:hep-ex/9910038 [hep-ex]} \BibitemShut {NoStop}%
\bibitem [{\citenamefont {Chekanov}\ \emph {et~al.}(2002)\citenamefont
  {Chekanov} \emph {et~al.}}]{Chekanov:2002xi}%
  \BibitemOpen
  \bibfield  {author} {\bibinfo {author} {\bibfnamefont {S.}~\bibnamefont
  {Chekanov}} \emph {et~al.} (\bibinfo {collaboration} {ZEUS}),\ }\href
  {\doibase 10.1007/s10052-002-0953-7} {\bibfield  {journal} {\bibinfo
  {journal} {Eur. Phys. J.}\ }\textbf {\bibinfo {volume} {C24}},\ \bibinfo
  {pages} {345} (\bibinfo {year} {2002})},\ \Eprint
  {http://arxiv.org/abs/hep-ex/0201043} {arXiv:hep-ex/0201043 [hep-ex]}
  \BibitemShut {NoStop}%
\bibitem [{\citenamefont {Chekanov}\ \emph {et~al.}(2003)\citenamefont
  {Chekanov} \emph {et~al.}}]{Chekanov:2002rm}%
  \BibitemOpen
  \bibfield  {author} {\bibinfo {author} {\bibfnamefont {S.}~\bibnamefont
  {Chekanov}} \emph {et~al.} (\bibinfo {collaboration} {ZEUS}),\ }\href
  {\doibase 10.1140/epjc/s2002-01079-0} {\bibfield  {journal} {\bibinfo
  {journal} {Eur. Phys. J.}\ }\textbf {\bibinfo {volume} {C26}},\ \bibinfo
  {pages} {389} (\bibinfo {year} {2003})},\ \Eprint
  {http://arxiv.org/abs/hep-ex/0205081} {arXiv:hep-ex/0205081 [hep-ex]}
  \BibitemShut {NoStop}%
\bibitem [{\citenamefont {Aktas}\ \emph {et~al.}(2003)\citenamefont {Aktas}
  \emph {et~al.}}]{Aktas:2003zi}%
  \BibitemOpen
  \bibfield  {author} {\bibinfo {author} {\bibfnamefont {A.}~\bibnamefont
  {Aktas}} \emph {et~al.} (\bibinfo {collaboration} {H1}),\ }\href {\doibase
  10.1016/j.physletb.2003.06.056} {\bibfield  {journal} {\bibinfo  {journal}
  {Phys. Lett.}\ }\textbf {\bibinfo {volume} {B568}},\ \bibinfo {pages} {205}
  (\bibinfo {year} {2003})},\ \Eprint {http://arxiv.org/abs/hep-ex/0306013}
  {arXiv:hep-ex/0306013 [hep-ex]} \BibitemShut {NoStop}%
\bibitem [{\citenamefont {Aktas}\ \emph {et~al.}(2006)\citenamefont {Aktas}
  \emph {et~al.}}]{Aktas:2005xu}%
  \BibitemOpen
  \bibfield  {author} {\bibinfo {author} {\bibfnamefont {A.}~\bibnamefont
  {Aktas}} \emph {et~al.} (\bibinfo {collaboration} {H1}),\ }\href {\doibase
  10.1140/epjc/s2006-02519-5} {\bibfield  {journal} {\bibinfo  {journal} {Eur.
  Phys. J.}\ }\textbf {\bibinfo {volume} {C46}},\ \bibinfo {pages} {585}
  (\bibinfo {year} {2006})},\ \Eprint {http://arxiv.org/abs/hep-ex/0510016}
  {arXiv:hep-ex/0510016 [hep-ex]} \BibitemShut {NoStop}%
\bibitem [{\citenamefont {Alexa}\ \emph {et~al.}(2013)\citenamefont {Alexa}
  \emph {et~al.}}]{Alexa:2013xxa}%
  \BibitemOpen
  \bibfield  {author} {\bibinfo {author} {\bibfnamefont {C.}~\bibnamefont
  {Alexa}} \emph {et~al.} (\bibinfo {collaboration} {H1}),\ }\href {\doibase
  10.1140/epjc/s10052-013-2466-y} {\bibfield  {journal} {\bibinfo  {journal}
  {Eur. Phys. J.}\ }\textbf {\bibinfo {volume} {C73}},\ \bibinfo {pages} {2466}
  (\bibinfo {year} {2013})},\ \Eprint {http://arxiv.org/abs/1304.5162}
  {arXiv:1304.5162 [hep-ex]} \BibitemShut {NoStop}%
\bibitem [{\citenamefont {Mäntysaari}\ and\ \citenamefont
  {Schenke}(2016{\natexlab{a}})}]{Mantysaari:2016ykx}%
  \BibitemOpen
  \bibfield  {author} {\bibinfo {author} {\bibfnamefont {H.}~\bibnamefont
  {Mäntysaari}}\ and\ \bibinfo {author} {\bibfnamefont {B.}~\bibnamefont
  {Schenke}},\ }\href {\doibase 10.1103/PhysRevLett.117.052301} {\bibfield
  {journal} {\bibinfo  {journal} {Phys. Rev. Lett.}\ }\textbf {\bibinfo
  {volume} {117}},\ \bibinfo {pages} {052301} (\bibinfo {year}
  {2016}{\natexlab{a}})},\ \Eprint {http://arxiv.org/abs/1603.04349}
  {arXiv:1603.04349 [hep-ph]} \BibitemShut {NoStop}%
\bibitem [{\citenamefont {Mäntysaari}\ and\ \citenamefont
  {Schenke}(2016{\natexlab{b}})}]{Mantysaari:2016jaz}%
  \BibitemOpen
  \bibfield  {author} {\bibinfo {author} {\bibfnamefont {H.}~\bibnamefont
  {Mäntysaari}}\ and\ \bibinfo {author} {\bibfnamefont {B.}~\bibnamefont
  {Schenke}},\ }\href {\doibase 10.1103/PhysRevD.94.034042} {\bibfield
  {journal} {\bibinfo  {journal} {Phys. Rev.}\ }\textbf {\bibinfo {volume}
  {D94}},\ \bibinfo {pages} {034042} (\bibinfo {year} {2016}{\natexlab{b}})},\
  \Eprint {http://arxiv.org/abs/1607.01711} {arXiv:1607.01711 [hep-ph]}
  \BibitemShut {NoStop}%
\bibitem [{\citenamefont {Schlichting}\ and\ \citenamefont
  {Schenke}(2014)}]{Schlichting:2014ipa}%
  \BibitemOpen
  \bibfield  {author} {\bibinfo {author} {\bibfnamefont {S.}~\bibnamefont
  {Schlichting}}\ and\ \bibinfo {author} {\bibfnamefont {B.}~\bibnamefont
  {Schenke}},\ }\href {\doibase 10.1016/j.physletb.2014.10.068} {\bibfield
  {journal} {\bibinfo  {journal} {Phys.Lett.}\ }\textbf {\bibinfo {volume}
  {B739}},\ \bibinfo {pages} {313} (\bibinfo {year} {2014})},\ \Eprint
  {http://arxiv.org/abs/1407.8458} {arXiv:1407.8458 [hep-ph]} \BibitemShut
  {NoStop}%
\bibitem [{\citenamefont {Albacete}\ and\ \citenamefont
  {Soto-Ontoso}(2016)}]{Albacete:2016pmp}%
  \BibitemOpen
  \bibfield  {author} {\bibinfo {author} {\bibfnamefont {J.~L.}\ \bibnamefont
  {Albacete}}\ and\ \bibinfo {author} {\bibfnamefont {A.}~\bibnamefont
  {Soto-Ontoso}},\ }\href@noop {} {\  (\bibinfo {year} {2016})},\ \Eprint
  {http://arxiv.org/abs/1605.09176} {arXiv:1605.09176 [hep-ph]} \BibitemShut
  {NoStop}%
\bibitem [{\citenamefont {Welsh}\ \emph {et~al.}(2016)\citenamefont {Welsh},
  \citenamefont {Singer},\ and\ \citenamefont {Heinz}}]{Welsh:2016siu}%
  \BibitemOpen
  \bibfield  {author} {\bibinfo {author} {\bibfnamefont {K.}~\bibnamefont
  {Welsh}}, \bibinfo {author} {\bibfnamefont {J.}~\bibnamefont {Singer}}, \
  and\ \bibinfo {author} {\bibfnamefont {U.~W.}\ \bibnamefont {Heinz}},\ }\href
  {\doibase 10.1103/PhysRevC.94.024919} {\bibfield  {journal} {\bibinfo
  {journal} {Phys. Rev.}\ }\textbf {\bibinfo {volume} {C94}},\ \bibinfo {pages}
  {024919} (\bibinfo {year} {2016})},\ \Eprint
  {http://arxiv.org/abs/1605.09418} {arXiv:1605.09418 [nucl-th]} \BibitemShut
  {NoStop}%
\bibitem [{\citenamefont {Cepila}\ \emph {et~al.}(2017)\citenamefont {Cepila},
  \citenamefont {Contreras},\ and\ \citenamefont
  {Tapia~Takaki}}]{Cepila:2016uku}%
  \BibitemOpen
  \bibfield  {author} {\bibinfo {author} {\bibfnamefont {J.}~\bibnamefont
  {Cepila}}, \bibinfo {author} {\bibfnamefont {J.~G.}\ \bibnamefont
  {Contreras}}, \ and\ \bibinfo {author} {\bibfnamefont {J.~D.}\ \bibnamefont
  {Tapia~Takaki}},\ }\href {\doibase 10.1016/j.physletb.2016.12.063} {\bibfield
   {journal} {\bibinfo  {journal} {Phys. Lett.}\ }\textbf {\bibinfo {volume}
  {B766}},\ \bibinfo {pages} {186} (\bibinfo {year} {2017})},\ \Eprint
  {http://arxiv.org/abs/1608.07559} {arXiv:1608.07559 [hep-ph]} \BibitemShut
  {NoStop}%
\bibitem [{\citenamefont {Mäntysaari}\ and\ \citenamefont
  {Schenke}(2017)}]{Mantysaari:2017dwh}%
  \BibitemOpen
  \bibfield  {author} {\bibinfo {author} {\bibfnamefont {H.}~\bibnamefont
  {Mäntysaari}}\ and\ \bibinfo {author} {\bibfnamefont {B.}~\bibnamefont
  {Schenke}},\ }\href@noop {} {\  (\bibinfo {year} {2017})},\ \Eprint
  {http://arxiv.org/abs/1703.09256} {arXiv:1703.09256 [hep-ph]} \BibitemShut
  {NoStop}%
\bibitem [{\citenamefont {Moreland}\ \emph {et~al.}(2017)\citenamefont
  {Moreland}, \citenamefont {Bernhard}, \citenamefont {Ke},\ and\ \citenamefont
  {Bass}}]{Moreland:2017kdx}%
  \BibitemOpen
  \bibfield  {author} {\bibinfo {author} {\bibfnamefont {J.~S.}\ \bibnamefont
  {Moreland}}, \bibinfo {author} {\bibfnamefont {J.~E.}\ \bibnamefont
  {Bernhard}}, \bibinfo {author} {\bibfnamefont {W.}~\bibnamefont {Ke}}, \ and\
  \bibinfo {author} {\bibfnamefont {S.~A.}\ \bibnamefont {Bass}},\ }in\
  \href@noop {} {\emph {\bibinfo {booktitle} {{}}}}\ (\bibinfo {year} {2017})\
  \Eprint {http://arxiv.org/abs/1704.04486} {arXiv:1704.04486 [nucl-th]}
  \BibitemShut {NoStop}%
\bibitem [{\citenamefont {Denicol}\ \emph {et~al.}(2016)\citenamefont
  {Denicol}, \citenamefont {Monnai},\ and\ \citenamefont
  {Schenke}}]{Denicol:2015nhu}%
  \BibitemOpen
  \bibfield  {author} {\bibinfo {author} {\bibfnamefont {G.}~\bibnamefont
  {Denicol}}, \bibinfo {author} {\bibfnamefont {A.}~\bibnamefont {Monnai}}, \
  and\ \bibinfo {author} {\bibfnamefont {B.}~\bibnamefont {Schenke}},\ }\href
  {\doibase 10.1103/PhysRevLett.116.212301} {\bibfield  {journal} {\bibinfo
  {journal} {Phys. Rev. Lett.}\ }\textbf {\bibinfo {volume} {116}},\ \bibinfo
  {pages} {212301} (\bibinfo {year} {2016})},\ \Eprint
  {http://arxiv.org/abs/1512.01538} {arXiv:1512.01538 [nucl-th]} \BibitemShut
  {NoStop}%
\bibitem [{\citenamefont {Schenke}\ \emph {et~al.}(2014)\citenamefont
  {Schenke}, \citenamefont {Tribedy},\ and\ \citenamefont
  {Venugopalan}}]{Schenke:2013dpa}%
  \BibitemOpen
  \bibfield  {author} {\bibinfo {author} {\bibfnamefont {B.}~\bibnamefont
  {Schenke}}, \bibinfo {author} {\bibfnamefont {P.}~\bibnamefont {Tribedy}}, \
  and\ \bibinfo {author} {\bibfnamefont {R.}~\bibnamefont {Venugopalan}},\
  }\href {\doibase 10.1103/PhysRevC.89.024901} {\bibfield  {journal} {\bibinfo
  {journal} {Phys. Rev.}\ }\textbf {\bibinfo {volume} {C89}},\ \bibinfo {pages}
  {024901} (\bibinfo {year} {2014})},\ \Eprint {http://arxiv.org/abs/1311.3636}
  {arXiv:1311.3636 [hep-ph]} \BibitemShut {NoStop}%
\bibitem [{\citenamefont {Huovinen}\ and\ \citenamefont
  {Petreczky}(2010)}]{Huovinen:2009yb}%
  \BibitemOpen
  \bibfield  {author} {\bibinfo {author} {\bibfnamefont {P.}~\bibnamefont
  {Huovinen}}\ and\ \bibinfo {author} {\bibfnamefont {P.}~\bibnamefont
  {Petreczky}},\ }\href {\doibase 10.1016/j.nuclphysa.2010.02.015} {\bibfield
  {journal} {\bibinfo  {journal} {Nucl. Phys.}\ }\textbf {\bibinfo {volume}
  {A837}},\ \bibinfo {pages} {26} (\bibinfo {year} {2010})},\ \Eprint
  {http://arxiv.org/abs/0912.2541} {arXiv:0912.2541 [hep-ph]} \BibitemShut
  {NoStop}%
\bibitem [{\citenamefont {Ryu}\ \emph {et~al.}(2015)\citenamefont {Ryu},
  \citenamefont {Paquet}, \citenamefont {Shen}, \citenamefont {Denicol},
  \citenamefont {Schenke}, \citenamefont {Jeon},\ and\ \citenamefont
  {Gale}}]{Ryu:2015vwa}%
  \BibitemOpen
  \bibfield  {author} {\bibinfo {author} {\bibfnamefont {S.}~\bibnamefont
  {Ryu}}, \bibinfo {author} {\bibfnamefont {J.~F.}\ \bibnamefont {Paquet}},
  \bibinfo {author} {\bibfnamefont {C.}~\bibnamefont {Shen}}, \bibinfo {author}
  {\bibfnamefont {G.~S.}\ \bibnamefont {Denicol}}, \bibinfo {author}
  {\bibfnamefont {B.}~\bibnamefont {Schenke}}, \bibinfo {author} {\bibfnamefont
  {S.}~\bibnamefont {Jeon}}, \ and\ \bibinfo {author} {\bibfnamefont
  {C.}~\bibnamefont {Gale}},\ }\href {\doibase 10.1103/PhysRevLett.115.132301}
  {\bibfield  {journal} {\bibinfo  {journal} {Phys. Rev. Lett.}\ }\textbf
  {\bibinfo {volume} {115}},\ \bibinfo {pages} {132301} (\bibinfo {year}
  {2015})},\ \Eprint {http://arxiv.org/abs/1502.01675} {arXiv:1502.01675
  [nucl-th]} \BibitemShut {NoStop}%
\bibitem [{\citenamefont {Jalilian-Marian}\ \emph {et~al.}(1997)\citenamefont
  {Jalilian-Marian}, \citenamefont {Kovner}, \citenamefont {Leonidov},\ and\
  \citenamefont {Weigert}}]{Jalilian-Marian:1997jx}%
  \BibitemOpen
  \bibfield  {author} {\bibinfo {author} {\bibfnamefont {J.}~\bibnamefont
  {Jalilian-Marian}}, \bibinfo {author} {\bibfnamefont {A.}~\bibnamefont
  {Kovner}}, \bibinfo {author} {\bibfnamefont {A.}~\bibnamefont {Leonidov}}, \
  and\ \bibinfo {author} {\bibfnamefont {H.}~\bibnamefont {Weigert}},\ }\href
  {\doibase 10.1016/S0550-3213(97)00440-9} {\bibfield  {journal} {\bibinfo
  {journal} {Nucl. Phys.}\ }\textbf {\bibinfo {volume} {B504}},\ \bibinfo
  {pages} {415} (\bibinfo {year} {1997})},\ \Eprint
  {http://arxiv.org/abs/hep-ph/9701284} {arXiv:hep-ph/9701284} \BibitemShut
  {NoStop}%
\bibitem [{\citenamefont {Jalilian-Marian}\ \emph {et~al.}(1999)\citenamefont
  {Jalilian-Marian}, \citenamefont {Kovner}, \citenamefont {Leonidov},\ and\
  \citenamefont {Weigert}}]{Jalilian-Marian:1997gr}%
  \BibitemOpen
  \bibfield  {author} {\bibinfo {author} {\bibfnamefont {J.}~\bibnamefont
  {Jalilian-Marian}}, \bibinfo {author} {\bibfnamefont {A.}~\bibnamefont
  {Kovner}}, \bibinfo {author} {\bibfnamefont {A.}~\bibnamefont {Leonidov}}, \
  and\ \bibinfo {author} {\bibfnamefont {H.}~\bibnamefont {Weigert}},\ }\href
  {\doibase 10.1103/PhysRevD.59.014014} {\bibfield  {journal} {\bibinfo
  {journal} {Phys. Rev.}\ }\textbf {\bibinfo {volume} {D59}},\ \bibinfo {pages}
  {014014} (\bibinfo {year} {1999})},\ \Eprint
  {http://arxiv.org/abs/hep-ph/9706377} {arXiv:hep-ph/9706377} \BibitemShut
  {NoStop}%
\bibitem [{\citenamefont {Iancu}\ \emph {et~al.}(2001)\citenamefont {Iancu},
  \citenamefont {Leonidov},\ and\ \citenamefont {McLerran}}]{Iancu:2000hn}%
  \BibitemOpen
  \bibfield  {author} {\bibinfo {author} {\bibfnamefont {E.}~\bibnamefont
  {Iancu}}, \bibinfo {author} {\bibfnamefont {A.}~\bibnamefont {Leonidov}}, \
  and\ \bibinfo {author} {\bibfnamefont {L.~D.}\ \bibnamefont {McLerran}},\
  }\href {\doibase 10.1016/S0375-9474(01)00642-X} {\bibfield  {journal}
  {\bibinfo  {journal} {Nucl. Phys.}\ }\textbf {\bibinfo {volume} {A692}},\
  \bibinfo {pages} {583} (\bibinfo {year} {2001})},\ \Eprint
  {http://arxiv.org/abs/hep-ph/0011241} {arXiv:hep-ph/0011241} \BibitemShut
  {NoStop}%
\bibitem [{\citenamefont {Denicol}\ \emph {et~al.}(2009)\citenamefont
  {Denicol}, \citenamefont {Kodama}, \citenamefont {Koide},\ and\ \citenamefont
  {Mota}}]{Denicol:2009am}%
  \BibitemOpen
  \bibfield  {author} {\bibinfo {author} {\bibfnamefont {G.}~\bibnamefont
  {Denicol}}, \bibinfo {author} {\bibfnamefont {T.}~\bibnamefont {Kodama}},
  \bibinfo {author} {\bibfnamefont {T.}~\bibnamefont {Koide}}, \ and\ \bibinfo
  {author} {\bibfnamefont {P.}~\bibnamefont {Mota}},\ }\href {\doibase
  10.1103/PhysRevC.80.064901} {\bibfield  {journal} {\bibinfo  {journal} {Phys.
  Rev.}\ }\textbf {\bibinfo {volume} {C80}},\ \bibinfo {pages} {064901}
  (\bibinfo {year} {2009})}\BibitemShut {NoStop}%
\bibitem [{\citenamefont {Rose}\ \emph {et~al.}(2014)\citenamefont {Rose},
  \citenamefont {Paquet}, \citenamefont {Denicol}, \citenamefont {Luzum},
  \citenamefont {Schenke}, \citenamefont {Jeon},\ and\ \citenamefont
  {Gale}}]{Rose:2014fba}%
  \BibitemOpen
  \bibfield  {author} {\bibinfo {author} {\bibfnamefont {J.-B.}\ \bibnamefont
  {Rose}}, \bibinfo {author} {\bibfnamefont {J.-F.}\ \bibnamefont {Paquet}},
  \bibinfo {author} {\bibfnamefont {G.~S.}\ \bibnamefont {Denicol}}, \bibinfo
  {author} {\bibfnamefont {M.}~\bibnamefont {Luzum}}, \bibinfo {author}
  {\bibfnamefont {B.}~\bibnamefont {Schenke}}, \bibinfo {author} {\bibfnamefont
  {S.}~\bibnamefont {Jeon}}, \ and\ \bibinfo {author} {\bibfnamefont
  {C.}~\bibnamefont {Gale}},\ }\bibfield  {booktitle} {\emph {\bibinfo
  {booktitle} {{}}},\ }\href {\doibase 10.1016/j.nuclphysa.2014.09.044}
  {\bibfield  {journal} {\bibinfo  {journal} {Nucl. Phys.}\ }\textbf {\bibinfo
  {volume} {A931}},\ \bibinfo {pages} {926} (\bibinfo {year} {2014})},\ \Eprint
  {http://arxiv.org/abs/1408.0024} {arXiv:1408.0024 [nucl-th]} \BibitemShut
  {NoStop}%
\bibitem [{\citenamefont {Rajagopal}\ and\ \citenamefont
  {Tripuraneni}(2010)}]{Rajagopal:2009yw}%
  \BibitemOpen
  \bibfield  {author} {\bibinfo {author} {\bibfnamefont {K.}~\bibnamefont
  {Rajagopal}}\ and\ \bibinfo {author} {\bibfnamefont {N.}~\bibnamefont
  {Tripuraneni}},\ }\href {\doibase 10.1007/JHEP03(2010)018} {\bibfield
  {journal} {\bibinfo  {journal} {JHEP}\ }\textbf {\bibinfo {volume} {1003}},\
  \bibinfo {pages} {018} (\bibinfo {year} {2010})}\BibitemShut {NoStop}%
\bibitem [{\citenamefont {Bhatt}\ \emph {et~al.}(2011)\citenamefont {Bhatt},
  \citenamefont {Mishra},\ and\ \citenamefont {Sreekanth}}]{Bhatt:2011kr}%
  \BibitemOpen
  \bibfield  {author} {\bibinfo {author} {\bibfnamefont {J.~R.}\ \bibnamefont
  {Bhatt}}, \bibinfo {author} {\bibfnamefont {H.}~\bibnamefont {Mishra}}, \
  and\ \bibinfo {author} {\bibfnamefont {V.}~\bibnamefont {Sreekanth}},\ }\href
  {\doibase 10.1016/j.physletb.2011.09.052} {\bibfield  {journal} {\bibinfo
  {journal} {Phys. Lett.}\ }\textbf {\bibinfo {volume} {B704}},\ \bibinfo
  {pages} {486} (\bibinfo {year} {2011})},\ \Eprint
  {http://arxiv.org/abs/1103.4333} {arXiv:1103.4333 [hep-ph]} \BibitemShut
  {NoStop}%
\bibitem [{\citenamefont {Habich}\ and\ \citenamefont
  {Romatschke}(2014)}]{Habich:2014tpa}%
  \BibitemOpen
  \bibfield  {author} {\bibinfo {author} {\bibfnamefont {M.}~\bibnamefont
  {Habich}}\ and\ \bibinfo {author} {\bibfnamefont {P.}~\bibnamefont
  {Romatschke}},\ }\href {\doibase 10.1007/JHEP12(2014)054} {\bibfield
  {journal} {\bibinfo  {journal} {JHEP}\ }\textbf {\bibinfo {volume} {12}},\
  \bibinfo {pages} {054} (\bibinfo {year} {2014})},\ \Eprint
  {http://arxiv.org/abs/1405.1978} {arXiv:1405.1978 [hep-ph]} \BibitemShut
  {NoStop}%
\bibitem [{\citenamefont {Denicol}\ \emph {et~al.}(2015)\citenamefont
  {Denicol}, \citenamefont {Gale},\ and\ \citenamefont
  {Jeon}}]{Denicol:2015bpa}%
  \BibitemOpen
  \bibfield  {author} {\bibinfo {author} {\bibfnamefont {G.~S.}\ \bibnamefont
  {Denicol}}, \bibinfo {author} {\bibfnamefont {C.}~\bibnamefont {Gale}}, \
  and\ \bibinfo {author} {\bibfnamefont {S.}~\bibnamefont {Jeon}},\ }\bibfield
  {booktitle} {\emph {\bibinfo {booktitle} {{}}},\ }\href@noop {} {\bibfield
  {journal} {\bibinfo  {journal} {PoS}\ }\textbf {\bibinfo {volume}
  {CPOD2014}},\ \bibinfo {pages} {033} (\bibinfo {year} {2015})},\ \Eprint
  {http://arxiv.org/abs/1503.00531} {arXiv:1503.00531 [nucl-th]} \BibitemShut
  {NoStop}%
\bibitem [{\citenamefont {Sanches}\ \emph {et~al.}(2015)\citenamefont
  {Sanches}, \citenamefont {Fogaça}, \citenamefont {Navarra},\ and\
  \citenamefont {Marrochio}}]{Sanches:2015vra}%
  \BibitemOpen
  \bibfield  {author} {\bibinfo {author} {\bibfnamefont {S.~M.}\ \bibnamefont
  {Sanches}}, \bibinfo {author} {\bibfnamefont {D.~A.}\ \bibnamefont
  {Fogaça}}, \bibinfo {author} {\bibfnamefont {F.~S.}\ \bibnamefont
  {Navarra}}, \ and\ \bibinfo {author} {\bibfnamefont {H.}~\bibnamefont
  {Marrochio}},\ }\href {\doibase 10.1103/PhysRevC.92.025204} {\bibfield
  {journal} {\bibinfo  {journal} {Phys. Rev.}\ }\textbf {\bibinfo {volume}
  {C92}},\ \bibinfo {pages} {025204} (\bibinfo {year} {2015})},\ \Eprint
  {http://arxiv.org/abs/1505.06335} {arXiv:1505.06335 [hep-ph]} \BibitemShut
  {NoStop}%
\bibitem [{\citenamefont {Nopoush}\ \emph {et~al.}(2014)\citenamefont
  {Nopoush}, \citenamefont {Ryblewski},\ and\ \citenamefont
  {Strickland}}]{Nopoush:2014pfa}%
  \BibitemOpen
  \bibfield  {author} {\bibinfo {author} {\bibfnamefont {M.}~\bibnamefont
  {Nopoush}}, \bibinfo {author} {\bibfnamefont {R.}~\bibnamefont {Ryblewski}},
  \ and\ \bibinfo {author} {\bibfnamefont {M.}~\bibnamefont {Strickland}},\
  }\href {\doibase 10.1103/PhysRevC.90.014908} {\bibfield  {journal} {\bibinfo
  {journal} {Phys. Rev.}\ }\textbf {\bibinfo {volume} {C90}},\ \bibinfo {pages}
  {014908} (\bibinfo {year} {2014})},\ \Eprint {http://arxiv.org/abs/1405.1355}
  {arXiv:1405.1355 [hep-ph]} \BibitemShut {NoStop}%
\bibitem [{\citenamefont {Alqahtani}\ \emph {et~al.}(2015)\citenamefont
  {Alqahtani}, \citenamefont {Nopoush},\ and\ \citenamefont
  {Strickland}}]{Alqahtani:2015qja}%
  \BibitemOpen
  \bibfield  {author} {\bibinfo {author} {\bibfnamefont {M.}~\bibnamefont
  {Alqahtani}}, \bibinfo {author} {\bibfnamefont {M.}~\bibnamefont {Nopoush}},
  \ and\ \bibinfo {author} {\bibfnamefont {M.}~\bibnamefont {Strickland}},\
  }\href {\doibase 10.1103/PhysRevC.92.054910} {\bibfield  {journal} {\bibinfo
  {journal} {Phys. Rev.}\ }\textbf {\bibinfo {volume} {C92}},\ \bibinfo {pages}
  {054910} (\bibinfo {year} {2015})},\ \Eprint
  {http://arxiv.org/abs/1509.02913} {arXiv:1509.02913 [hep-ph]} \BibitemShut
  {NoStop}%
\bibitem [{\citenamefont {Shen}\ \emph {et~al.}(2016)\citenamefont {Shen},
  \citenamefont {Qiu}, \citenamefont {Song}, \citenamefont {Bernhard},
  \citenamefont {Bass},\ and\ \citenamefont {Heinz}}]{Shen:2014vra}%
  \BibitemOpen
  \bibfield  {author} {\bibinfo {author} {\bibfnamefont {C.}~\bibnamefont
  {Shen}}, \bibinfo {author} {\bibfnamefont {Z.}~\bibnamefont {Qiu}}, \bibinfo
  {author} {\bibfnamefont {H.}~\bibnamefont {Song}}, \bibinfo {author}
  {\bibfnamefont {J.}~\bibnamefont {Bernhard}}, \bibinfo {author}
  {\bibfnamefont {S.}~\bibnamefont {Bass}}, \ and\ \bibinfo {author}
  {\bibfnamefont {U.}~\bibnamefont {Heinz}},\ }\href {\doibase
  10.1016/j.cpc.2015.08.039} {\bibfield  {journal} {\bibinfo  {journal}
  {Comput. Phys. Commun.}\ }\textbf {\bibinfo {volume} {199}},\ \bibinfo
  {pages} {61} (\bibinfo {year} {2016})},\ \Eprint
  {http://arxiv.org/abs/1409.8164} {arXiv:1409.8164 [nucl-th]} \BibitemShut
  {NoStop}%
\bibitem [{\citenamefont {Abelev}\ \emph {et~al.}(2014)\citenamefont {Abelev}
  \emph {et~al.}}]{Abelev:2013haa}%
  \BibitemOpen
  \bibfield  {author} {\bibinfo {author} {\bibfnamefont {B.~B.}\ \bibnamefont
  {Abelev}} \emph {et~al.} (\bibinfo {collaboration} {ALICE}),\ }\href
  {\doibase 10.1016/j.physletb.2013.11.020} {\bibfield  {journal} {\bibinfo
  {journal} {Phys. Lett.}\ }\textbf {\bibinfo {volume} {B728}},\ \bibinfo
  {pages} {25} (\bibinfo {year} {2014})},\ \Eprint
  {http://arxiv.org/abs/1307.6796} {arXiv:1307.6796 [nucl-ex]} \BibitemShut
  {NoStop}%
\bibitem [{\citenamefont {Chatrchyan}\ \emph
  {et~al.}(2013{\natexlab{b}})\citenamefont {Chatrchyan} \emph
  {et~al.}}]{Chatrchyan:2013nka}%
  \BibitemOpen
  \bibfield  {author} {\bibinfo {author} {\bibfnamefont {S.}~\bibnamefont
  {Chatrchyan}} \emph {et~al.} (\bibinfo {collaboration} {CMS Collaboration}),\
  }\href {\doibase 10.1016/j.physletb.2013.06.028} {\bibfield  {journal}
  {\bibinfo  {journal} {Phys.Lett.}\ }\textbf {\bibinfo {volume} {B724}},\
  \bibinfo {pages} {213} (\bibinfo {year} {2013}{\natexlab{b}})},\ \Eprint
  {http://arxiv.org/abs/1305.0609} {arXiv:1305.0609 [nucl-ex]} \BibitemShut
  {NoStop}%
\bibitem [{\citenamefont {Aad}\ \emph {et~al.}(2014)\citenamefont {Aad} \emph
  {et~al.}}]{Aad:2014lta}%
  \BibitemOpen
  \bibfield  {author} {\bibinfo {author} {\bibfnamefont {G.}~\bibnamefont
  {Aad}} \emph {et~al.} (\bibinfo {collaboration} {ATLAS}),\ }\href {\doibase
  10.1103/PhysRevC.90.044906} {\bibfield  {journal} {\bibinfo  {journal} {Phys.
  Rev.}\ }\textbf {\bibinfo {volume} {C90}},\ \bibinfo {pages} {044906}
  (\bibinfo {year} {2014})},\ \Eprint {http://arxiv.org/abs/1409.1792}
  {arXiv:1409.1792 [hep-ex]} \BibitemShut {NoStop}%
\bibitem [{\citenamefont {Abelev}\ \emph
  {et~al.}(2013{\natexlab{b}})\citenamefont {Abelev} \emph
  {et~al.}}]{ABELEV:2013wsa}%
  \BibitemOpen
  \bibfield  {author} {\bibinfo {author} {\bibfnamefont {B.~B.}\ \bibnamefont
  {Abelev}} \emph {et~al.} (\bibinfo {collaboration} {ALICE Collaboration}),\
  }\href {\doibase 10.1016/j.physletb.2013.08.024} {\bibfield  {journal}
  {\bibinfo  {journal} {Phys.Lett.}\ }\textbf {\bibinfo {volume} {B726}},\
  \bibinfo {pages} {164} (\bibinfo {year} {2013}{\natexlab{b}})},\ \Eprint
  {http://arxiv.org/abs/1307.3237} {arXiv:1307.3237 [nucl-ex]} \BibitemShut
  {NoStop}%
\bibitem [{\citenamefont {Khachatryan}\ \emph {et~al.}(2015)\citenamefont
  {Khachatryan} \emph {et~al.}}]{Khachatryan:2014jra}%
  \BibitemOpen
  \bibfield  {author} {\bibinfo {author} {\bibfnamefont {V.}~\bibnamefont
  {Khachatryan}} \emph {et~al.} (\bibinfo {collaboration} {CMS}),\ }\href
  {\doibase 10.1016/j.physletb.2015.01.034} {\bibfield  {journal} {\bibinfo
  {journal} {Phys. Lett.}\ }\textbf {\bibinfo {volume} {B742}},\ \bibinfo
  {pages} {200} (\bibinfo {year} {2015})},\ \Eprint
  {http://arxiv.org/abs/1409.3392} {arXiv:1409.3392 [nucl-ex]} \BibitemShut
  {NoStop}%
\bibitem [{\citenamefont {Chojnacki}\ \emph {et~al.}(2012)\citenamefont
  {Chojnacki}, \citenamefont {Kisiel}, \citenamefont {Florkowski},\ and\
  \citenamefont {Broniowski}}]{Chojnacki:2011hb}%
  \BibitemOpen
  \bibfield  {author} {\bibinfo {author} {\bibfnamefont {M.}~\bibnamefont
  {Chojnacki}}, \bibinfo {author} {\bibfnamefont {A.}~\bibnamefont {Kisiel}},
  \bibinfo {author} {\bibfnamefont {W.}~\bibnamefont {Florkowski}}, \ and\
  \bibinfo {author} {\bibfnamefont {W.}~\bibnamefont {Broniowski}},\ }\href
  {\doibase 10.1016/j.cpc.2011.11.018} {\bibfield  {journal} {\bibinfo
  {journal} {Comput. Phys. Commun.}\ }\textbf {\bibinfo {volume} {183}},\
  \bibinfo {pages} {746} (\bibinfo {year} {2012})},\ \Eprint
  {http://arxiv.org/abs/1102.0273} {arXiv:1102.0273 [nucl-th]} \BibitemShut
  {NoStop}%
\bibitem [{\citenamefont {Pratt}(1986)}]{Pratt:1986cc}%
  \BibitemOpen
  \bibfield  {author} {\bibinfo {author} {\bibfnamefont {S.}~\bibnamefont
  {Pratt}},\ }\href {\doibase 10.1103/PhysRevD.33.1314} {\bibfield  {journal}
  {\bibinfo  {journal} {Phys. Rev.}\ }\textbf {\bibinfo {volume} {D33}},\
  \bibinfo {pages} {1314} (\bibinfo {year} {1986})}\BibitemShut {NoStop}%
\bibitem [{\citenamefont {Bertsch}\ \emph {et~al.}(1988)\citenamefont
  {Bertsch}, \citenamefont {Gong},\ and\ \citenamefont
  {Tohyama}}]{Bertsch:1988db}%
  \BibitemOpen
  \bibfield  {author} {\bibinfo {author} {\bibfnamefont {G.}~\bibnamefont
  {Bertsch}}, \bibinfo {author} {\bibfnamefont {M.}~\bibnamefont {Gong}}, \
  and\ \bibinfo {author} {\bibfnamefont {M.}~\bibnamefont {Tohyama}},\ }\href
  {\doibase 10.1103/PhysRevC.37.1896} {\bibfield  {journal} {\bibinfo
  {journal} {Phys. Rev.}\ }\textbf {\bibinfo {volume} {C37}},\ \bibinfo {pages}
  {1896} (\bibinfo {year} {1988})}\BibitemShut {NoStop}%
\bibitem [{\citenamefont {Plumberg}\ and\ \citenamefont
  {Heinz}(2016)}]{Plumberg:2016sig}%
  \BibitemOpen
  \bibfield  {author} {\bibinfo {author} {\bibfnamefont {C.}~\bibnamefont
  {Plumberg}}\ and\ \bibinfo {author} {\bibfnamefont {U.}~\bibnamefont
  {Heinz}},\ }\href@noop {} {\  (\bibinfo {year} {2016})},\ \Eprint
  {http://arxiv.org/abs/1611.03161} {arXiv:1611.03161 [nucl-th]} \BibitemShut
  {NoStop}%
\bibitem [{\citenamefont {Adam}\ \emph {et~al.}(2015)\citenamefont {Adam} \emph
  {et~al.}}]{Adam:2015pya}%
  \BibitemOpen
  \bibfield  {author} {\bibinfo {author} {\bibfnamefont {J.}~\bibnamefont
  {Adam}} \emph {et~al.} (\bibinfo {collaboration} {ALICE}),\ }\href {\doibase
  10.1103/PhysRevC.91.034906} {\bibfield  {journal} {\bibinfo  {journal} {Phys.
  Rev.}\ }\textbf {\bibinfo {volume} {C91}},\ \bibinfo {pages} {034906}
  (\bibinfo {year} {2015})},\ \Eprint {http://arxiv.org/abs/1502.00559}
  {arXiv:1502.00559 [nucl-ex]} \BibitemShut {NoStop}%
\bibitem [{\citenamefont {Shapoval}\ \emph {et~al.}(2013)\citenamefont
  {Shapoval}, \citenamefont {Braun-Munzinger}, \citenamefont {Karpenko},\ and\
  \citenamefont {Sinyukov}}]{Shapoval:2013jca}%
  \BibitemOpen
  \bibfield  {author} {\bibinfo {author} {\bibfnamefont {V.~M.}\ \bibnamefont
  {Shapoval}}, \bibinfo {author} {\bibfnamefont {P.}~\bibnamefont
  {Braun-Munzinger}}, \bibinfo {author} {\bibfnamefont {I.~A.}\ \bibnamefont
  {Karpenko}}, \ and\ \bibinfo {author} {\bibfnamefont {{\relax Yu}.~M.}\
  \bibnamefont {Sinyukov}},\ }\href {\doibase 10.1016/j.physletb.2013.07.002}
  {\bibfield  {journal} {\bibinfo  {journal} {Phys. Lett.}\ }\textbf {\bibinfo
  {volume} {B725}},\ \bibinfo {pages} {139} (\bibinfo {year} {2013})},\ \Eprint
  {http://arxiv.org/abs/1304.3815} {arXiv:1304.3815 [hep-ph]} \BibitemShut
  {NoStop}%
\bibitem [{\citenamefont {Keegan}\ \emph
  {et~al.}(2016{\natexlab{b}})\citenamefont {Keegan}, \citenamefont {Kurkela},
  \citenamefont {Mazeliauskas},\ and\ \citenamefont {Teaney}}]{Keegan:2016cpi}%
  \BibitemOpen
  \bibfield  {author} {\bibinfo {author} {\bibfnamefont {L.}~\bibnamefont
  {Keegan}}, \bibinfo {author} {\bibfnamefont {A.}~\bibnamefont {Kurkela}},
  \bibinfo {author} {\bibfnamefont {A.}~\bibnamefont {Mazeliauskas}}, \ and\
  \bibinfo {author} {\bibfnamefont {D.}~\bibnamefont {Teaney}},\ }\href
  {\doibase 10.1007/JHEP08(2016)171} {\bibfield  {journal} {\bibinfo  {journal}
  {JHEP}\ }\textbf {\bibinfo {volume} {08}},\ \bibinfo {pages} {171} (\bibinfo
  {year} {2016}{\natexlab{b}})},\ \Eprint {http://arxiv.org/abs/1605.04287}
  {arXiv:1605.04287 [hep-ph]} \BibitemShut {NoStop}%
\bibitem [{\citenamefont {Schenke}\ and\ \citenamefont
  {Schlichting}(2016)}]{Schenke:2016ksl}%
  \BibitemOpen
  \bibfield  {author} {\bibinfo {author} {\bibfnamefont {B.}~\bibnamefont
  {Schenke}}\ and\ \bibinfo {author} {\bibfnamefont {S.}~\bibnamefont
  {Schlichting}},\ }\href {\doibase 10.1103/PhysRevC.94.044907} {\bibfield
  {journal} {\bibinfo  {journal} {Phys. Rev.}\ }\textbf {\bibinfo {volume}
  {C94}},\ \bibinfo {pages} {044907} (\bibinfo {year} {2016})},\ \Eprint
  {http://arxiv.org/abs/1605.07158} {arXiv:1605.07158 [hep-ph]} \BibitemShut
  {NoStop}%
\bibitem [{\citenamefont {Jia}\ \emph {et~al.}(2017)\citenamefont {Jia},
  \citenamefont {Zhou},\ and\ \citenamefont {Trzupek}}]{Jia:2017hbm}%
  \BibitemOpen
  \bibfield  {author} {\bibinfo {author} {\bibfnamefont {J.}~\bibnamefont
  {Jia}}, \bibinfo {author} {\bibfnamefont {M.}~\bibnamefont {Zhou}}, \ and\
  \bibinfo {author} {\bibfnamefont {A.}~\bibnamefont {Trzupek}},\ }\href@noop
  {} {\  (\bibinfo {year} {2017})},\ \Eprint {http://arxiv.org/abs/1701.03830}
  {arXiv:1701.03830 [nucl-th]} \BibitemShut {NoStop}%
\end{thebibliography}%

\end{document}